## Abstract

*Purpose*

Cardiac substructure delineation is emerging in treatment planning to minimize the risk of radiation-induced heart disease. Deep learning offers efficient methods to reduce contouring burden but currently lacks generalizability across different modalities and overlapping structures. This work introduces and validates a Modality-AGnostic Image Cascade (MAGIC) deep-learning pipeline for comprehensive and multi-modal cardiac substructure segmentation.

*Materials and methods*

MAGIC is implemented through replicated encoding and decoding branches of an nnU-Net backbone to handle multi-modality inputs and overlapping labels. First benchmarked on the multi-modality whole-heart segmentation (MMWHS) dataset including cardiac CT-angiography (CCTA) and MR modalities, twenty cardiac substructures (heart, chambers, great vessels (GVs), valves, coronary arteries (CAs), and conduction nodes) from clinical simulation CT (Sim-CT), low-field MR-Linac, and cardiac CT-angiography (CCTA) modalities were delineated to train semi-supervised (n=151), validate (n=15), and test (n=30) MAGIC. For comparison, fourteen single-modality comparison models (two MMWHS modalities and four subgroups across three clinical modalities) were trained. Methods were evaluated for efficiency and against reference contours through the Dice similarity coefficient (DSC) and two-tailed Wilcoxon Signed-Rank test ($p<0.05$).

Results

Average MMWHS DSC scores across CCTA and MR inputs were 0.88±0.08 and 0.87±0.04 respectively with significant improvement ($p<0.05$) over unimodal baselines. Average 20-structure DSC scores were 0.75±0.16 (heart, 0.96±0.01; chambers, 0.89±0.05; GVs, 0.81±0.09; CAs, 0.60±0.13; valves, 0.70±0.18; nodes, 0.66±0.12) for Sim-CT, 0.68±0.21 (heart, 0.94±0.01; chambers, 0.87±0.05; GVs, 0.72±0.18; CAs, 0.50±0.18; valves, 0.62±0.16; nodes, 0.52±0.16) for MR-Linac, and 0.80±0.16 (heart, 0.95±0.01; chambers, 0.93±0.04; GVs, 0.84±0.06; CAs, 0.77±0.12; valves, 0.68±0.23; nodes, 0.72±0.11) for CCTA. Furthermore, >80% and >70% reductions in training time and parameters were achieved, respectively.

*Conclusions*

MAGIC offers an efficient, lightweight solution capable of segmenting multiple image modalities and overlapping structures in a single model without compromising segmentation accuracy.

**Conflicts of Interest**

C.K.G-H reports research collaborations with RaySearch Laboratories, Medscint, and Leo Cancer Care outside of the submitted work (PI: Carri Glide-Hurst). S.Z. reports grants or contracts from Google and Radiation Oncology Institute; consulting fees from Radformation, LEK Consulting; 3 pending patents on AI for radiation treatment planning; and is a founder of intelliOnc LLC. N.S. reports travel funding from American Association of Physicists in Medicine and the Wisconsin Scholarship Hub. P.N. reports grants or contracts from GE Healthcare; Consulting feeds from Canon Medical and WCG Clinical; Board member of Canon Medical; and owns stock or stock options in Moderna; Q.H., A.K., C.R., J.P., A.I.G., A.K., J.Z., and M.D. report no conflicts of interest.

**Author contributions**

**Nicholas Summerfield:** Conceptualization, Data Curation, Formal Analysis, Investigation, Methodology, Resources, Software, Validation, Visualization, Writing – original draft, Writing – review & editing. **Qisheng He:** Conceptualization, Investigation, Methodology, Software, Writing – review & editing. **Alex Kuo:** Data Curation, Writing – review & editing. **Chase Ruff:** Data Curation, Writing – review & editing. **Joshua Pan:** Methodology, Software, Writing – review & editing. **Ahmed I. Ghanem:** Data Curation, Writing – review & editing. **Simeng Zhu:** Data Curation, Writing – review & editing. **Anudeep Kumar:** Methodology, Software, Writing – review & editing. **Prashant Nagpal:** Supervision, Writing – review & editing. **Jiwei Zhao:** Formal Analysis, Writing – review & editing. **Ming Dong:** Conceptualization, Methodology, Project Administration, Supervision, Writing – review & editing. **Carri Glide-Hurst:** Conceptualization, Funding acquisition, Methodology, Project Administration, Supervision, Writing – original draft, Writing – review & editing.

## Manuscript

### Introduction

Radiation-induced heart disease is a known complication of thoracic radiation therapy (RT) encompassing both acute and late effects[1,2]. The current standard of care is to delineate the whole heart (WH) as a single organ at risk and use simple metrics to evaluate dose-to-volume relationships for potential cardiac risk[3]. However, radiation dose to cardiac substructures has been more strongly associated with cardiotoxicities[1] and cardiac substructures have been proposed as sparing structures in advanced cardiac sparing RT plans[4,5].

Deep learning-based segmentation for cardiac substructures have offered efficient solutions to implement for RT[6]. Yet, at present, these methods are unimodal, targeted to a single, specific image type such as computed tomography (CT) or magnetic resonance (MR) imaging and challenged by modality-specific, specialized contouring and lengthy delineation times[7]. RT has historically been performed using CT[6,8–12]. However, with MR-guided RT[13] (MR coupled with linear accelerators, MR-Linac) and the availability of cardiac gating technology (e.g., cardiac CT-angiography (CCTA)), multiple imaging modalities can now be used to evaluate cardiovascular anatomy[6]. While CCTA is primarily a diagnostic imaging method, similar acquisitions can be obtained through intravenous contrast enhancement[9,10] and cardiac-gated Sim-CT[14]. With various anatomical variations for coronary arteries, contrast based CCTA offers a great opportunity to accurately accommodate potential overlaps with adjoining cardiac substructures and may potentially offer a clinical benefit. CCTA has been implemented to aid structure delineation[15] and has recently been recommended to help guide RT planning[16], albeit not a part of standard-of-care yet. Recent methods target a specific modality, are incapable of generalizing to different inputs without large performance loss or prediction failure[17,18], and limit the advantages of different contrasts and resolutions available between modalities. Morris et al.[12] incorporated paired Sim-CT and T2-weighted MR images during training to aid in model fitting and boost accuracy. However, clinical use remained unimodal in application to Sim-CT inputs. At present, separate modality-specific, unimodal models have been required for each treatment and imaging modality available in RT[17,18].

Cardiovascular anatomy is complex and generally considered to encompass the WH, chambers, great vessels (GVs), coronary arteries (CAs), valves, and the conduction nodes[19–22]. Nodes have recently been proposed as an additional sparing structure due to strong connections with conduction disorders[22]. When combined into a composite label, several structures overlap with more than one true label per image voxel[19,20] (e.g. WH, chambers, valves, and atrioventricular node may overlap, resulting in four potentially correct identifications for a single voxel). Semantic segmentation by definition is incapable of handling overlapping labels whereas realizing each structure's full volume is essential to calculating substructure-specific dosimetric endpoints in RT planning[23]. Methods presented by recent publications[8–11] employ specifically trained models to tackle subgroups of non-overlapping cardiac substructures. These methods require multiple trained models and inferences whereas a single model, capable of segmenting a composite mask of cardiac substructures in a single inference, to our knowledge, has yet to be investigated.

Methods proposed to handle overlapping structures in literature include hierarchical segmentation through topological and region-based optimization as well as multi-task segmentation. Topological methods introduce containment/exclusion constraints for nested structures[24,25]. Region-based optimization partitions structures into subsections of overlap/non-overlap as different classifications that are reconstructed at output[26]. Multi-task segmentation uses multi-head or multi-branch decoding layers to separate the outputs of a single model into individual semantic segmentation tasks[27]. Cardiac anatomy[19,20] contains densely overlapping structures that are often not strictly nested, contain ambiguous interfaces, and is subject to large class imbalances[28] that may worsen through a fine-grained segmentation hierarchy. As an alternative, multi-task segmentation offers an efficient solution enabling groups of non-overlapping structures to be individually targeted.

Due to overlapping conditions, at least four models would be necessary to automatically segment cardiac substructures currently of interest in the literature[8–11], most commonly one model per modality. Maintaining numerous models presents a bottleneck for clinical adoption where institutional-specific implementation and re-occurring model updates to address data drift are common[29,30]. Thus, there is a clinical need for an efficient segmentation algorithm, capable of handling both overlapping structures and multi-modal data in a single model. To address these current limitations, we present a Modality-AGnostic Image Cascade (MAGIC) framework, benchmark it on the Multi-Modality Whole Heart Segmentation (MMWHS) dataset[31], and validate it on a clinical dataset for three modalities and 20 cardiac substructures, thereby laying the technical groundwork and generalizability for future prospective modality-agnostic clinical trials.

## Methods and Materials

## Benchmark Cohort

The MMWHS datasets[31] consists of 40 unpaired CCTA and cardiac MR images (n=20 per modality) from different sites and scanners. On each image seven cardiac substructures are identified including the chambers (left/right atrium/ventricle (LA, RA, LV, RV) and myocardium (Myo)), ascending aorta (AA), and pulmonary artery (PA).

## Multi-Institutional, Clinical Cohort

The multi-institutional, clinical dataset includes three imaging modalities that were retrospectively retrieved in this institutional review board-approved study from two different institutions: Sim-CT (62 patients, n=65 unique datasets), MR-Linac (34 patients, 2-3 fractionated images per patient used for training augmentation[13], n=66 unique datasets), and CCTA (n=65 patients and datasets). Both MR-Linac and Sim-CT images were obtained from patients undergoing RT at one of two institutions for either thoracic or abdominal cancers adjacent to the heart. Diagnostic CCTA images were collected from a single institution following standard clinical protocols, with a field-of-view encompassing the heart.

Non-contrast Sim-CTs were acquired on a Brilliance Big Bore CT simulator (Philips Medical Systems, Cleveland, OH) or a Somatom Edge CT scanner (Siemens Healthineers, Malvern, PA) with 0.98-1.4x0.98-1.4x2-3mm$^3$ resolution, using 120-140kVp, and 100-434mAs. To facilitate generalizability to many treatment strategies, CT data included free breathing (n=22), deep-inspiration breath-hold (n=1), average 4DCT (n=36) or 50% end-exhalation of a 4DCT (n=6) matching clinical treatment scenarios. CCTAs were acquired with 60-100ml of intravenous contrast, 80-140kVp, automated mAs following institutional protocols, and underwent axial reconstruction with $0.49x0.49x0.63mm^3$ resolution following a vendor-provided deep-learning algorithm (SnapShot Freeze 2, GE Healthcare) to minimize cardiac motion. MR-Linac images were acquired on a 0.35T MR-Linac (ViewRay Systems, Denver, CO) at both institutions using clinically available T1/T2 balanced steady state free precession sequences (TrueFISP, Siemens, MAGNETOM Avanto, Syngo MR B19) with 17-25 seconds scan time, 1.5-$1.6x1.5\text{-}1.6x3mm^3$ resolution, 60° flip angle, and phase encoding in anterior-poster plane.

Images were manually labeled with 20 cardiac substructures including WH; chambers (LA, RA, LV, RV); GVs (AA, superior/inferior vena cava (SVC, IVC), PA, and pulmonary veins (PVs)); CAs (right (RCA), left main (LMCA), left anterior descending (LADA), and left circumflex (LCx) arteries); valves (aortic (V-AV), pulmonic (V-PV), mitral (V-MV), and tricuspid (V-TV) valves); and conduction nodes (sinoatrial (N-SA) and atrioventricular (N-AV) nodes) following a consensus of published guidelines[19–22]. Expert consultation was performed with a radiologist with cardiovascular subspecialty and 10+ years of experience.

## MAGIC Methods

To facilitate multi-modal semantic segmentation of overlapping structures, we developed the Modality-AGnostic Image Cascade (MAGIC), as shown in Figure 1A. Modality-agnostic/cross-modality segmentation has been demonstrated previously through statistic pooling[32], region-aware fusion[33], intra-modal transformers[34], and separate normalization layers[35]. MAGIC adopts an approach supported by our previous work[17,18] using modality-specific encoding branches connected at the bottleneck (i.e. lowest resolution connection between encoding and decoding branches[36]) to learn and encode each modality to a common latent space. Overlapping structures are handled through parallel multi-decoder branches split after the bottleneck (Table A1), each targeting a group of non-overlapping structures in a unified model. By optimizing the shared branches with all modalities, MAGIC learns from a large source of diverse features across modalities which has been shown to improve model robustness[17,18] while remaining generalized to use on multiple different inputs. Similarly, by optimizing all segmentation groups with each encoding branch, cross-correlation between structures is preserved, further improving prediction confidence[37]. During training, a single optimizer is attached to all branches to harmonize learning and preserve gradient momentum across modalities[38] while only updating weights of encoding branches corresponding to the input. At each iteration, an image is randomly selected from all available datasets to avoid biasing the model towards any single modality.

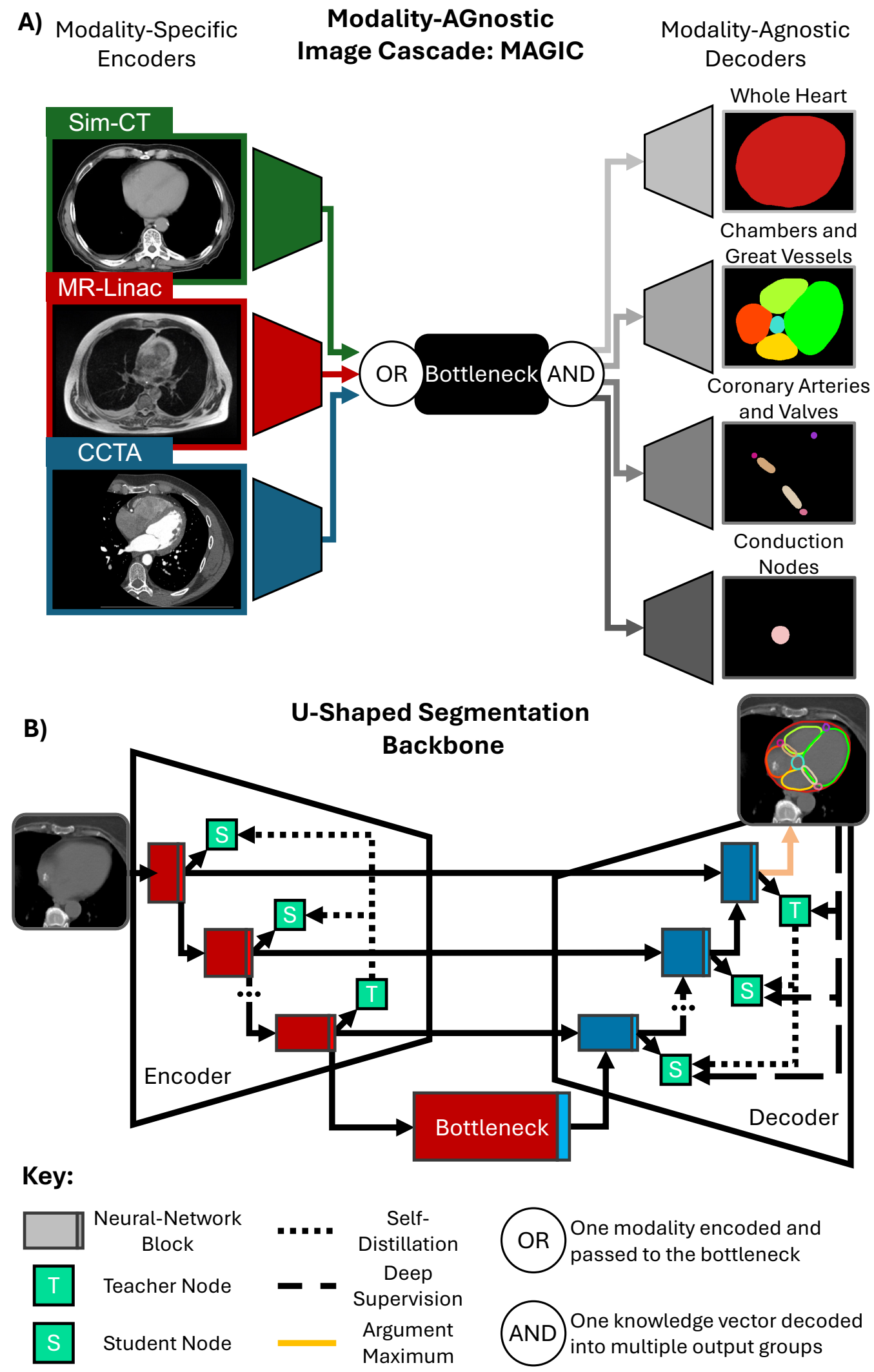

Figure 1: Modality-AGnostic Image Cascade (MAGIC, A, top) and U-shaped segmentation backbone (B, bottom). The modality-agnostic and multi-label capabilities of MAGIC arise from duplicated modality-specific encoders and modality-agnostic decoders connected to a single bottleneck.

The MAGIC framework can be applied to any U-shaped network, as shown in Figure 1B, by replicating encoding and decoding branches to match the dataset. Optimization and architectural designs—including skip connections, deep supervision, and self-distillation—remain applicable across the MAGIC framework. Building upon our prior work, we implemented MAGIC with a modified patch-based nnU-Net with dual self-distillation (nnU-Net.wSD) that was optimized through the nnU-Net's pipeline[39] and previously validated on the MMWHS dataset[31,36] and a clinical dataset of 12 cardiac substructures on MR-Linac images[13]. MAGIC configured on MMWHS consists of two encoding branches for CCTA and MR inputs. Whereas for the clinical dataset, three encoder branches for MR-Linac, Sim-CT, and CCTA images, and four decoder branches were initialized for four subgroups of cardiac substructures: A) WH; B) chambers and GVs; C) CAs and valves; D) conduction nodes. To compare MAGIC to conventional approaches, fourteen "unimodal" models (two modalities for MMWHS experiments and four segmentation subgroups across three modalities for the clinical experiments) were trained[8–11] with equivalent hyperparameters and training protocols as their MAGIC counterpart. This implementation of MAGIC has been made publicly available on GitHub[1].

**Training Methods**

All work was conducted in Python (version 3.10) using the open-source packages PyTorch[40] (version 2.6) and MONAI[41] (version 0.9), with training and evaluation performed on an Nvidia A4000 graphics processing unit with 48 GB of memory. The MMWHS dataset was split into 70% for training, 10% for validation, and 20% for testing[35]. For the clinical cohort, manually labeled datasets for each modality were split for testing (n=10), validation (n=5), and training (Sim-CT and CCTA, n=25; MR-Linac, n=26) to train a fully supervised MAGIC. This model was used to predict contours on a cohort of unlabeled images where an additional 25 images per modality were visually assessed for accuracy and included without alteration in semi-supervised retraining of re-initialized MAGIC. The same cohort of reference and predicted labels were used in training all unimodal models under equivalent model training configurations.

To improve generalizability[42], all images underwent standardized pre-processing steps and training augmentations. Optimization was performed over 500 epochs and constrained by a hybrid loss function, guided through a combined Dice, Dice-boundary, and focal loss as described in the Appendix. The Dice loss addresses the label imbalance inherent to segmentation tasks[43], the Dice-boundary loss applies emphasis on the surface shape of the structure[44], and the focal loss aids in label classification tuned to the segmentation task of each group[28,43]. The nnU-Net.wSD backbone model utilizes additional losses to facilitate dual self-distillation during optimization[36] which are combined linearly with the primary loss. After each epoch, the hold-out validation set was evaluated to identify the best model weights based on the Dice similarity coefficient (DSC) averaged across all substructures and modalities. Additional details on preprocessing and training steps can be found in Appendix A.

---

[1] github.com/NRSummerfield/MAGIC

**Evaluation techniques**

MAGIC and unimodal segmentation frameworks were compared for model efficiency through comparing the time for training and inference, training memory, and the quantity of trainable parameters on the clinical dataset. Performance was measured quantitatively[23,45] through DSC, 95% Hausdorff Distance (HD95), and mean absolute surface distance (MASD) using MONAI[41] (version 0.9) and normalized surface distance[2] (NSD; tolerance, 3mm; alternatively known as surface DSC). Due to their thin, tube-like geometries, CA evaluation is challenged as small deviations largely impact volumetric metrics such as DSC. Therefore, the CAs were further evaluated using the centerline DSC[3] (clDice), a novel metric previously used for brain and retina vessel segmentation[45,46]. Quantitative metrics comparing MAGIC and unimodal models were assessed through two-tailed Wilcoxon signed-rank tests with $p<0.05$ considered significantly different. If a model fails to predict a structure, HD95 and MASD were replaced with the image diagonal[45] and a two-tailed Mann-Whitney U test was used to assess statistical differences ($p<0.05$). Finally, predicted and reference contours from the clinical dataset were visually compared against underlying anatomy by a trained cardiology annotator for global performance characteristics.

---

[2] *github.com/google-deepmind/surface-distance*

[3] *github.com/jocpae/clDice*

## Results

For the clinical experiments, the unimodal models required a total of 185 hours, maximally 14GB of memory per model, and 736 million parameters to train. In contrast, MAGIC required only 31 hours (~83% reduction), 194 million parameters (~74% reduction), and 32GB of memory. During inference, the unimodal models required a total (time summed over the four models required to segment all 20 structures) of 15.3±0.01 seconds to generate a composite label whereas MAGIC required 11.4±0.01 seconds in a single inference. Graphical loss and validation history presented in Figure B1.

**MMWHS Evaluation**

| | | DSC (AU) | | | NSD (AU; τ = 3 mm) | | | HD95 (mm) | | | MASD (mm) | | |
|---|---|---|---|---|---|---|---|---|---|---|---|---|---|
| | | MAGIC | Unimodal | p<0.05 | MAGIC | Unimodal | p<0.05 | MAGIC | Unimodal | p<0.05 | MAGIC | Unimodal | p<0.05 |
| CCTA | Average | 0.88 ± 0.08 | 0.85 ± 0.13 | * | 0.91 ± 0.09 | 0.85 ± 0.15 | * | 8.9 ± 12.0 | 11.7 ± 13.4 | * | 2.2 ± 2.4 | 3.2 ± 4.9 | * |
| | Chambers | 0.87 ± 0.09 | 0.83 ± 0.14 | * | 0.90 ± 0.10 | 0.82 ± 0.16 | * | 6.4 ± 7.9 | 11.9 ± 13.7 | * | 1.9 ± 1.8 | 3.6 ± 5.6 | * |
| | GVs | 0.91 ± 0.07 | 0.89 ± 0.07 | | 0.91 ± 0.09 | 0.91 ± 0.08 | | 15.1 ± 17.1 | 11.2 ± 12.9 | | 3.1 ± 3.4 | 2.1 ± 2.4 | |
| MR | Average | 0.87 ± 0.04 | 0.84 ± 0.07 | * | 0.89 ± 0.06 | 0.86 ± 0.09 | * | 11.0 ± 14.3 | 13.4 ± 16.4 | * | 1.7 ± 1.1 | 2.1 ± 2.0 | * |
| | Chambers | 0.88 ± 0.04 | 0.86 ± 0.07 | * | 0.90 ± 0.06 | 0.87 ± 0.09 | * | 5.4 ± 2.1 | 8.4 ± 7.9 | * | 1.5 ± 0.5 | 2.0 ± 1.9 | * |
| | GVs | 0.83 ± 0.03 | 0.81 ± 0.05 | | 0.86 ± 0.05 | 0.84 ± 0.07 | | 24.9 ± 20.8 | 26.1 ± 23.7 | | 2.4 ± 1.7 | 2.5 ± 2.2 | |
| Combined | Average | 0.87 ± 0.07 | 0.85 ± 0.10 | * | 0.90 ± 0.08 | 0.85 ± 0.12 | * | 9.9 ± 13.2 | 12.6 ± 15.0 | * | 2.0 ± 1.9 | 2.6 ± 3.8 | * |
| | Chambers | 0.88 ± 0.07 | 0.84 ± 0.11 | * | 0.90 ± 0.08 | 0.85 ± 0.13 | * | 5.9 ± 5.8 | 10.2 ± 11.3 | * | 1.7 ± 1.3 | 2.8 ± 4.3 | * |
| | Myo | 0.84 ± 0.06 | 0.79 ± 0.07 | * | 0.95 ± 0.03 | 0.91 ± 0.06 | * | 3.6 ± 1.0 | 5.9 ± 3.9 | * | 1.2 ± 0.3 | 1.4 ± 0.2 | * |
| | LA | 0.85 ± 0.10 | 0.78 ± 0.19 | * | 0.84 ± 0.11 | 0.75 ± 0.22 | * | 8.1 ± 5.8 | 14.3 ± 17.2 | * | 1.8 ± 0.8 | 5.2 ± 8.3 | * |
| | LV | 0.93 ± 0.02 | 0.91 ± 0.03 | * | 0.96 ± 0.03 | 0.91 ± 0.04 | * | 3.5 ± 0.8 | 5.7 ± 2.4 | | 1.2 ± 0.4 | 1.5 ± 0.4 | * |
| | RA | 0.87 ± 0.05 | 0.86 ± 0.04 | | 0.87 ± 0.07 | 0.84 ± 0.06 | * | 9.5 ± 9.9 | 9.3 ± 7.4 | | 2.6 ± 2.5 | 2.4 ± 1.6 | |
| | RV | 0.90 ± 0.03 | 0.86 ± 0.08 | * | 0.90 ± 0.07 | 0.83 ± 0.12 | * | 4.9 ± 1.9 | 15.5 ± 13.6 | * | 1.5 ± 0.4 | 3.4 ± 2.9 | * |
| | GVs | 0.87 ± 0.06 | 0.85 ± 0.07 | | 0.89 ± 0.07 | 0.88 ± 0.08 | | 20.0 ± 19.6 | 18.6 ± 20.5 | | 2.8 ± 2.7 | 2.3 ± 2.3 | |
| | AA | 0.88 ± 0.08 | 0.88 ± 0.09 | | 0.89 ± 0.08 | 0.89 ± 0.10 | | 21.1 ± 21.2 | 16.7 ± 21.0 | | 2.3 ± 2.7 | 1.1 ± 0.6 | |
| | PA | 0.85 ± 0.04 | 0.83 ± 0.05 | * | 0.88 ± 0.06 | 0.86 ± 0.06 | * | 18.9 ± 17.9 | 20.6 ± 19.8 | | 3.3 ± 2.6 | 3.5 ± 2.7 | |

Table 1: Quantitative performance of MAGIC and unimodal comparator models on the Multi-Modality Whole-Heart Segmentation (MMWHS)[31] challenge dataset for Dice similarity coefficient (DSC), normalized surface distance (NSD; tolerance, 3mm), 95% Hausdorff distance, (HD95) and mean absolute surface distance (MASD) on CCTA (top), MR (middle), and all combined (bottom) inputs. Abbreviations: Arbitrary units, AU; rest defined in text.

Quantitative comparisons of MAGIC against two unimodal models (one for CCTA, the other for MRI) on the MMWHS dataset are shown in Table 1 and Table B1. Implementing MAGIC significantly improved (p<0.05) the average DSC, NSD, HD95, and MASD for both CCTA and MR inputs over equivalent unimodal models.

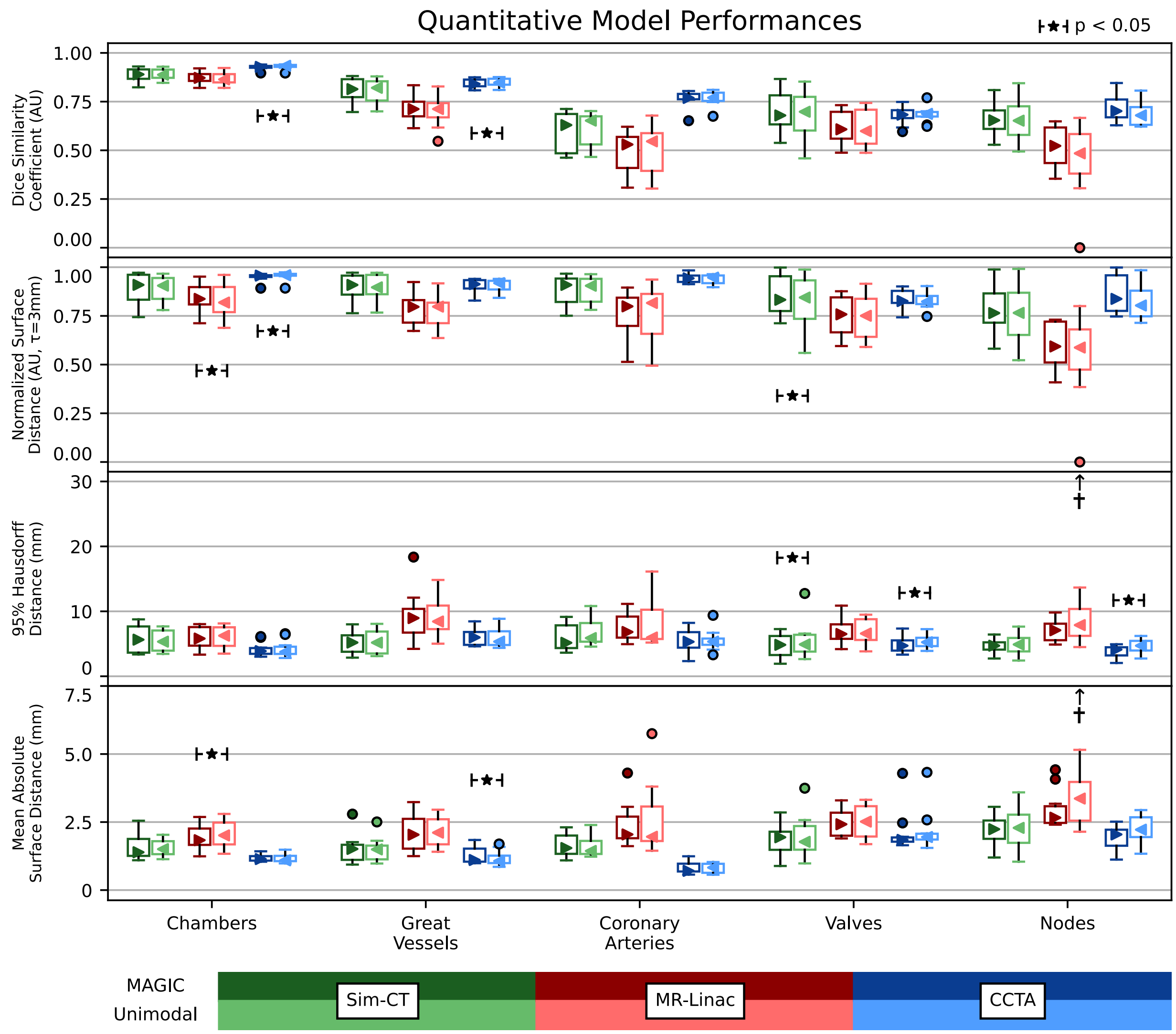

Figure 2: Quantitative model performance comparing MAGIC against 12 specialized comparator models examining the Dice similarity coefficient, normalized surface distance (tolerance, 3mm), 95% Hausdorff distance (HD95), and mean absolute surface distance (MASD) for Sim-CT, MR-Linac, and CCTA inputs. Abbreviations: Arbitrary Units, AU.

† comparator model failed to predict an atrioventricular node resulting in infinite metrics for HD95 and MASD and was replaced with the image diagonal for analysis.

Quantitative comparisons between MAGIC and equivalently trained unimodal models on the clinical dataset are shown in Figure 2 with substructure-specific performance comparisons presented in Tables B2-5. MAGIC conserved strong overall agreement with reference contours across all three modalities and WH predictions >0.94 DSC (results not shown). Average DSC scores across all 20 structures were 0.75±0.16 (NSD, 0.87±0.13; HD95, 5.3±2.8mm; MASD, 1.7±0.8mm) for Sim-CT, 0.68±0.21 (NSD, 0.77±0.18; HD95, 7.3±4.8mm; MASD, 2.3±1.2mm) for MR-Linac, and 0.80±0.16 (NSD, 0.90±0.14; HD95, 5.1±3.6mm; MASD, 1.4±1.2mm) for CCTA images. The unimodal model for MR-Linac cardiac node segmentation failed to predict N-AV for one of the test cases whereas MAGIC yielded an average prediction (DSC, 0.50; HD95, 7.5mm). MAGIC outperformed single-modality comparators in 53%, 73%, and 46% of metric comparisons for Sim-CT, MR-Linac, and CCTA respectively. Overall, MAGIC's performance was comparable to that of 12 unimodal models (four segmentation subgroups across three modalities) with limited statistical differences across comparisons.

A subset of qualitative axial comparisons and volumetric renderings are shown in Figure 3 for the clinical dataset. Best, average, and worst-performing cases were selected based on combined substructure-average DSC. The best cases demonstrated strong consistency with the reference contours, with discrepancies mainly observed at the superior/inferior extents. The worst-performing Sim-CT case was for a 50% gated 4DCT image that had increased noise and lower soft-tissue contrast compared to the best and average cases, potentially contributing to reduced performance. Both MAGIC and unimodal models were robust to MR-Linac artifacts (i.e. average and worst-case MR-

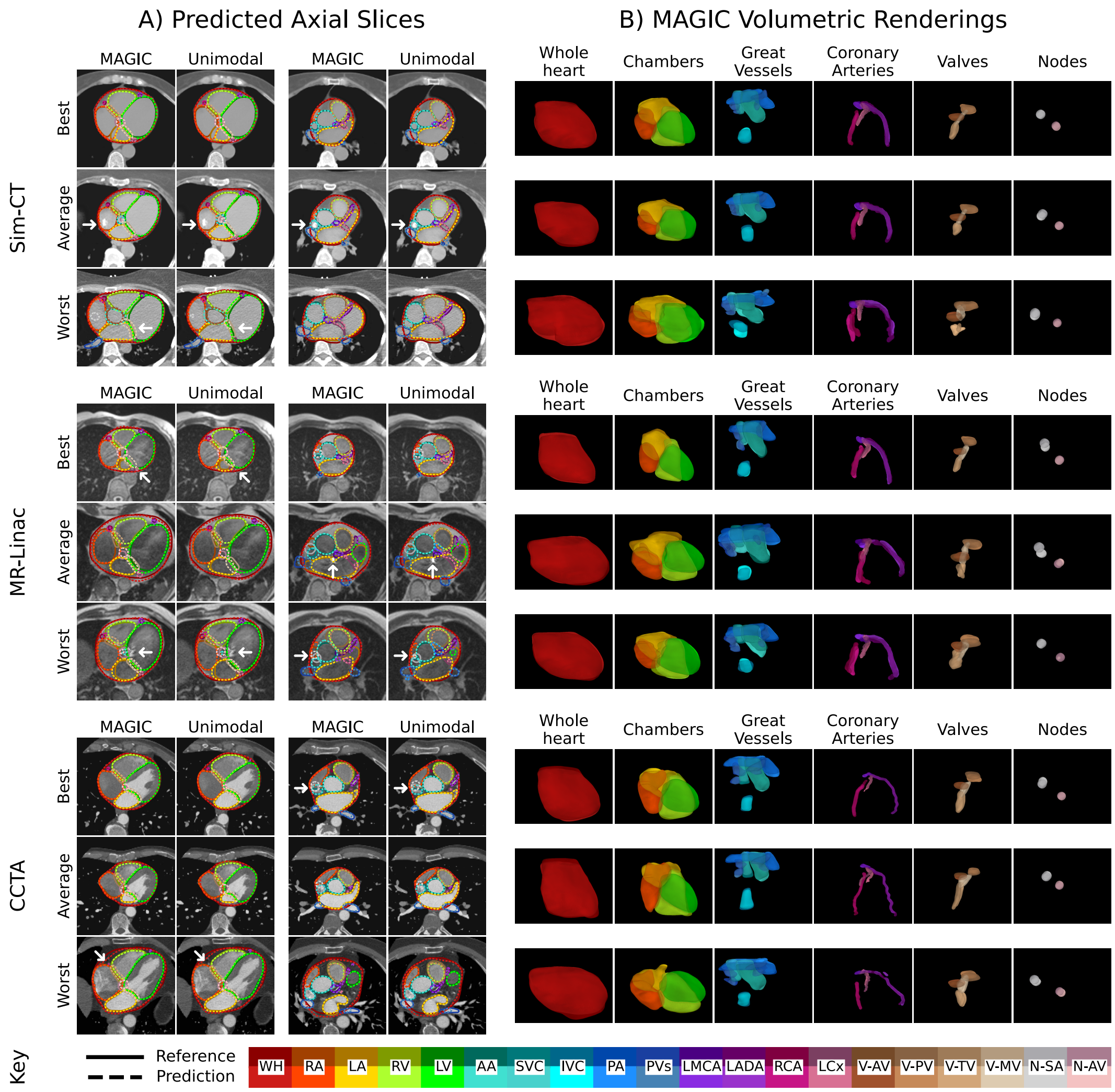

Figure 3: Qualitative analysis highlighting best, average, and worst patient-average Dice similarity coefficient across Sim-CT (top), MR-Linac (middle), and CCTA (bottom) inputs. A) Axial slices comparing MAGIC and unimodal predictions against reference contours with arrows identifying differences in model performance and performance in the presence of image artifacts or patient implants. B) Volumetric renderings of MAGIC predictions overlayed with reference contours. Abbreviations defined in text.

Linac, Figure 3). Finally, all CCTA cases show strong visual agreement, with the primary areas of disagreement arising from conflicting coronary veins or sinus that visually resemble CAs in some scans.

Detailed CA analysis is presented in Figure 4 for visual comparisons of MAGIC predicted centerlines against reference contours. Average clDice for MAGIC was 0.84±0.19 for Sim-CT, 0.61±0.34 for MR-Linac, and 0.87±0.16 for CCTA with no significant difference compared to unimodal methods. For each modality, the best-case predictions achieved clDice scores >0.93 with centerlines that fall nearly completely within the reference contours, while closely matching the tract and length of the reference in the average cases. Greater deviations in posterior-inferior regions were present in the worst cases for some RCA and LCx predictions whereas LADA predictions consistently had clDice scores >0.80 showing high agreement even in the worst cases.

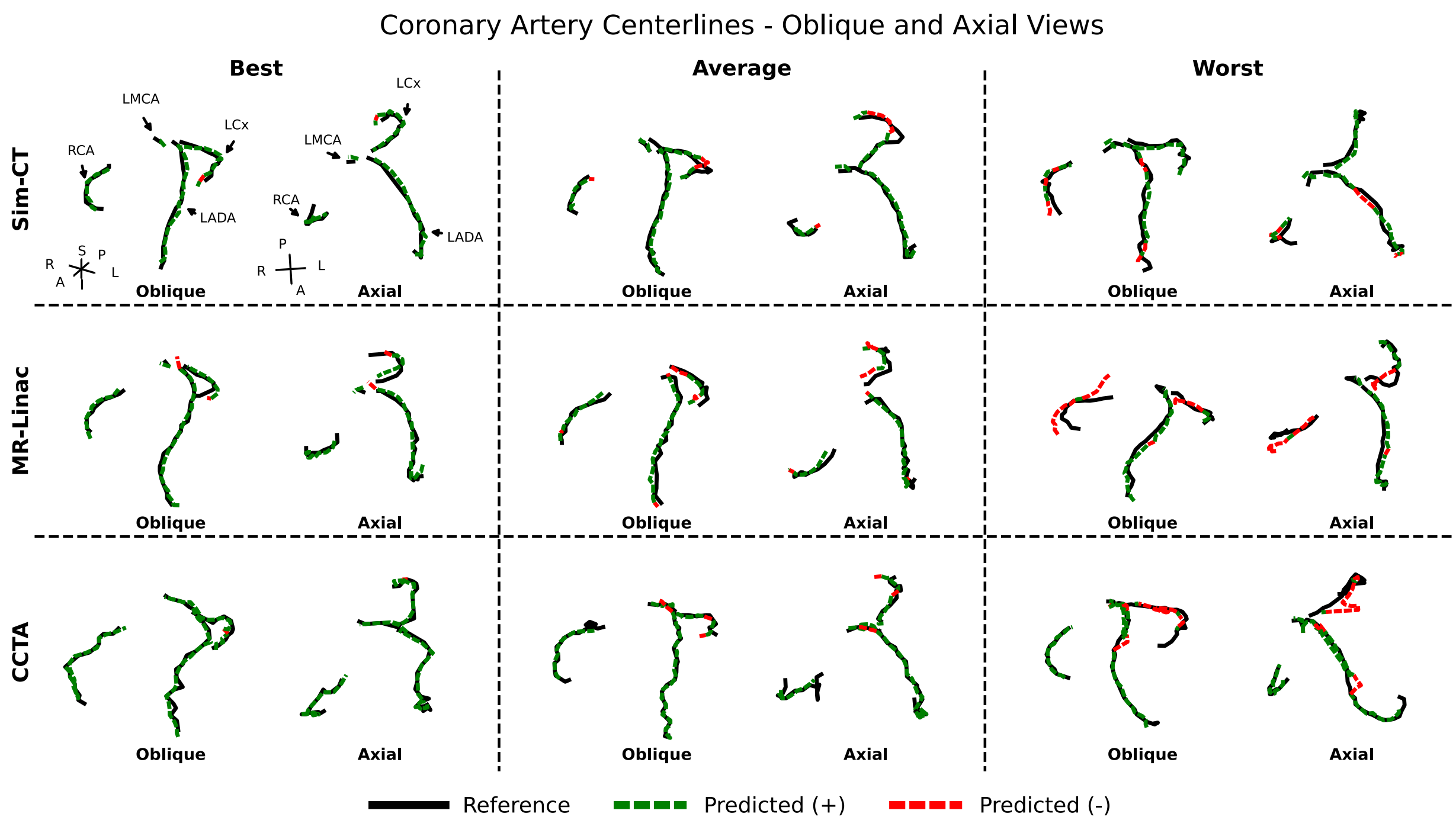

Figure 4: Centerline review of MAGIC predictions against reference contours for best, average, and worst patient-average clDice. Positive (+, green) predictions correspond with the centerline remaining within the reference contour and negative (-, red) predictions deviate outside. Abbreviations defined in text.

## Discussion

This work introduced and validated MAGIC, a deep-learning framework applicable to multiple imaging modalities and capable of segmenting overlapping cardiac substructures in a single, lightweight model. By incorporating multiple modality-specific encoders coupled with a shared modality-agnostic bottleneck and decoder branches, MAGIC offers a flexible solution for universal segmentation while remaining compatible with U-shaped semantic segmentation backbones and advanced optimization strategies. When benchmarked on the MMWHS dataset, MAGIC significantly improved performance on both inputs. For clinical models implemented on Sim-CT, MR-Linac, and CCTA images, MAGIC performed comparably to 12 state-of-the-art unimodal counterparts for cardiac substructure segmentation, yet offers a substantially simplified approach, achieving >80% training time reduction, faster inference speeds, and requiring >70% fewer parameters.

| Structure Groups | Dice Similarity Coefficient (AU) | | | | | | |
|---|---|---|---|---|---|---|---|
| | Sim-CT | | | | | Low-Field MR-Linac | |
| | MAGIC | Chen et al.[8] | Momin et al.[9] | Harms et al[10] | Van der Pole et al.[11] | MAGIC | Summerfield et al.[13] |
| Whole Heart | 0.96 | 0.95 | 0.97 | 0.96 | - | 0.94 | - |
| Chambers | 0.89 | 0.91 | 0.94 | 0.94 | Median: 0.89 | 0.87 | 0.85 |
| Great Vessels | 0.81 | 0.86 | 0.92 | 0.93 | Median: 0.78 | 0.72 | 0.67 |
| Coronary Arteries | 0.60 | 0.60 | 0.70 | 0.66 | Median: 0.34 | 0.50 | 0.33 |
| Valves | 0.70 | 0.81 | 0.74 | 0.77 | Median: 0.35 | 0.62 | - |
| Nodes | 0.66 | - | - | - | - | 0.52 | - |

| Structure Groups | 95% Hausdorff Distance (mm) | | | | | | |
|---|---|---|---|---|---|---|---|
| | Sim-CT | | | | | Low-Field MR-Linac | |
| | MAGIC | Chen et al.[8] | Momin et al.[9] | Harms et al[10] | Van der Pole et al.[11] | MAGIC | Summerfield et al.[13] |
| Whole Heart | 4.3 | 4.1 | 5.6 | 6.0 | - | 5.3 | - |
| Chambers | 5.8 | 4.4 | 6.7 | 4.6 | Median: 6.5 | 5.8 | 7.0 |
| Great Vessels | 5.3 | 6.0 | 5.4 | 4.4 | Median: 11.1 | 9.2 | 9.9 |
| Coronary Arteries | 6.0 | 7.3 | 6.4 | 7.2 | Median: 20.7 | 7.6 | 9.5 |
| Valves | 4.7 | 4.4 | 12.6 | 5.4 | Median: 11.6 | 6.8 | - |
| Nodes | 4.6 | - | - | - | - | 7.0 | - |

Table 2: Average model performance comparing MAGIC to state-of-the-art methods[4,8–10,13] on Sim-CT and low-field MR-Linac inputs. Abbreviations: Arbitrary Units, AU.

Comparisons between MAGIC and other state-of-the-art methods are presented in Table 2 for Sim-CT and MR-Linac inputs where values are directly reference from literature[8–11,13]. Chen et al.[8] proposed implementing four nnU-Nets to segment 19 cardiac substructures. Our single MAGIC performed comparably, demonstrating more accurate HD95, but slightly lower DSCs, potentially due to different in-plane spatial resolution (<1mm compared to 1.5mm) and a larger modality-specific training dataset (n=75 vs n=25 reference contours). Both Momin et al.[9] and Harms et al.[10] used multi-stage networks with 3-4 independent modules to segment 16 cardiac substructures from a mixed set of non-contrast and contrast-enhanced Sim-CT images. Here, presented methods outperform MAGIC in DSC, likely due to the use of mixed contrast scans, yet MAGIC remains comparable with HD95 performance. Van der Pol et al.[11] used two nnU-Nets to segment 16 cardiac substructure for a cohort of ventricular tachycardia (VT) patients. MAGIC offered superior DSC and HD95 with similar Sim-CT imaging resolution. However, performance differences likely arose from the presence of implanted cardiac devices in the VT cohort. Finally, MAGIC's performance on Sim-CT was compared against TotalSegmentator[47] with details provided in appendix B.

We previously performed cardiac substructure segmentation on low-field MR-Linac images using a single nnU-Net.wSD on 12 non-overlapping structures[13]. Comparatively, MAGIC offered improved segmentation performance with added contours for the WH, valves, and conduction nodes. Between the presented modalities, MAGIC had the lowest performance on MR-Linac data, possibly due to lower image resolution compared to CT (MR-Linac, 1.5-1.6x1.5-1.6x3mm$^3$; Sim-CT, 0.98-1.4x0.98-1.4x2-3mm$^3$) and the presence of artifacts[13].

Diagnostic CCTA has been utilized for cardiac segmentation, but limited comparisons exist for composite labels of multiple substructures. The MMWHS challenge[31] was targeted to chambers and GVs wherein the best method yielded an average DSC of 0.91 and 100% Hausdorff distance (HD100) of 25.2mm. On comparable structures, MAGIC has an equivalent DSC of 0.91 but far superior HD100 of 9.5mm (recalculated; HD95, 4.9mm). Gharleghi et al.[48] reported a CA challenge where all CAs were considered a single classification. The best method yielded a DSC of 0.87 and HD95 of 4.2mm. Comparatively, MAGIC classified four unique CAs, achieving an average DSC of 0.77 and HD95 of 5.4mm. Decreased DSC is likely due to MAGIC being targeted to a more complex multi-class segmentation with additional label uncertainty[49]. A limitation of MAGIC on the clinical dataset is that CCTA inputs were down sampled to match the lower-resolution RT volumes. Due to the high quality of CCTA imaging and intravenous contrast, key information is conserved and enables MAGIC to predict all 20 structures with average DSC > 0.80 and the highest accuracy of the included modalities.

For CAs, MAGIC yielded an average clDice[46] of 0.84, 0.61, and 0.87 for Sim-CT, MR-Linac, and CCTA respectively, with no statistical differences from unimodal comparators. Performance was lowest for LMCA on MR-Linac with an average of 0.45 but maintained an HD95 of 6.3mm. This lower performance is likely due to disagreements in the superior/inferior direction in addition to the short length of the LMCA[20]. On Sim-CT and CCTA, structure averages range between 0.79-0.96, suggesting strong overall predictions for the CAs. The LCx had the most uncertainty as seen on the worst MR-Linac and CCTA cases. Visual inspection of predicted centerlines demonstrated strong agreement with little to no deviations of predicted centerlines outside of reference contours for most structures.

To our knowledge, MAGIC is the first deep-learning model that includes the conduction nodes. Loap et al.[50] utilized a multi-atlas segmentation, achieving a median DSC of 0.56 and 0.15 for the N-SA and N-AV respectively. Finnegan et al.[51] proposed a geometric based approach, achieving an average DSC of 0.47 (HD95, 8.9mm) and 0.46 (HD95, 8.4mm) for the N-SA, and N-AV respectively. In comparison, MAGIC scored a DSC of 0.65 (HD95, 4.8mm) and 0.67 (HD95, 4.4mm) for N-SA and N-AV respectively on Sim-CT. While no comparator data are available, MAGIC showed promising performance on both MR-Linac (DSC, 0.52; HD95, 7.0mm) and CCTA images (DSC, 0.72; HD95, 3.8mm).

In this paper, MAGIC has been benchmarked on MMWHS[31], compared against TotalSegmentator[47], and underwent ablation of the implemented multi-branch decoder. In our prior work[17,18,36], the algorithms supporting MAGIC were further extensively evaluated on public datasets, outperformed other state-of-the-art algorithms, and incorporated physician review of cardiac segmentation[13] and were therefore not pursued in this current body of work. While addressing cross-modality segmentation, Dou et al.[35] presented a single, lightweight U-Net with modality-specific normalization layers and shared-parameter convolutions for both MR and CT segmentation. This could be seamlessly integrated with MAGIC by conjoining the encoding branches and preserving the split-decoders from multi-task learning. While MAGIC handles overlapping structures through multi-branch decoders[27], alternative approaches may include hierarchical, region-based segmentation that has shown promise for brain tumor segmentation[26], vessel wall segmentation[24], and cardiac ventricle segmentation[25]. Future work may include evaluation on clinical acceptability and dosimetric comparisons[8] which were beyond the scope of this study on model development.

In addition to large efficiency gains (>80% reduced training time and >70% reduced model parameters), MAGIC yielded ~2mm reductions of MASD and HD95 for MR-Linac predictions. Furthermore, MAGIC significantly improved both CCTA and MR performance when evaluated on the MMWHS challenge dataset. Finally, MAGIC remained comparable on Sim-CT and CCTA inputs for the clinical experiments when compared against equivalently trained, unimodal baselines. This potentially demonstrates benefits of cross-modality learning as documented in literature[17,18,52]. Natural extensions include appending the encoders to include other images such as dual-energy and calcium-scoring CT, high-field MR-Linac, diagnostic MR for cardiac segmentation, and incorporate paired patient imaging where additional benefits of cross-modality learning may be further realized.

Despite these limitations, MAGIC remains generalizable to multiple different inputs while predicting 20 composite cardiac substructures in a single inference and has been validated on clinical Sim-CT, MR-Linac, and CCTA images. MAGIC produces competitive predictions against state-of-the-art unimodal models, reducing training time by >80% with >70% fewer parameters. This flexibility reduces burden for institutions adopting multi-modal imaging pipelines, functional imaging, different RT platforms, and adaptive workflows[7,18]. Beyond cardiac applications, MAGIC may benefit other disease sites such as head and neck[53] or pelvis[54], where multi-contrast imaging and overlapping structures are common. With imaging diversity growing across disciplines, the built-in adaptability of MAGIC introduces clinically practicable solutions. This work will further be employed for a future prospective clinical trial to support modality-agnostic treatment strategies.

**References:**

1. Chin V, Finnegan RN, Keall P, Otton J, Delaney GP, Vinod SK. Overview of cardiac toxicity from radiation therapy. *J Med Imaging Radiat Oncol*. 2024;68(8):987-1000. doi:10.1111/1754-9485.13757
2. Walls GM, Bergom C, Mitchell JD, et al. Cardiotoxicity following thoracic radiotherapy for lung cancer. *Br J Cancer*. 2025;132(4):311-325. doi:10.1038/s41416-024-02888-0
3. Gagliardi G, Constine LS, Moiseenko V, et al. Radiation dose-volume effects in the heart. *Radiation Oncology Biology*. 2010;76:S77-S85. doi:10.1016/j.ijrobp.2009.04.093
4. van der Pol LHG, Hackett SL, Hoesein FAAM, et al. On the feasibility of cardiac substructure sparing in magnetic resonance imaging guided stereotactic lung radiotherapy. *Med Phys*. 2023;50(1):397-409. doi:10.1002/mp.16028
5. Morris ED, Aldridge K, Ghanem AI, Zhu S, Glide-Hurst CK. Incorporating sensitive cardiac substructure sparing into radiation therapy planning. *J Appl Clin Med Phys*. 2020;21(11):195-204. doi:10.1002/acm2.13037
6. Finnegan RN, Quinn A, Booth J, et al. Cardiac substructure delineation in radiation therapy – A state-of-the-art review. *J Med Imaging Radiat Oncol*. Published online May 17, 2024. doi:10.1111/1754-9485.13668
7. Chen C, Qin C, Qiu H, et al. Deep Learning for Cardiac Image Segmentation: A Review. *Frontiers in Cardiovascular Medicine | www.frontiersin.org*. 2020;7:25. doi:10.3389/fcvm.2020.00025
8. Chen X, Mumme RP, Corrigan KL, et al. Deep learning–based automatic segmentation of cardiac substructures for lung cancers. *Radiotherapy and Oncology*. Published online December 2023:110061. doi:10.1016/j.radonc.2023.110061
9. Momin S, Lei Y, McCall NS, et al. Mutual enhancing learning-based automatic segmentation of CT cardiac substructure. *Phys Med Biol*. 2022;67(10). doi:10.1088/1361-6560/ac692d
10. Harms J, Lei Y, Tian S, et al. Automatic delineation of cardiac substructures using a region-based fully convolutional network. *Med Phys*. 2021;48(6):2867-2876. doi:10.1002/mp.14810
11. van der Pol LHG, Blanck O, Grehn M, et al. Auto-contouring of cardiac substructures for Stereotactic arrhythmia radioablation (STAR): A STOPSTORM.eu consortium study. *Radiotherapy and Oncology*. 2025;202:110610. doi:10.1016/j.radonc.2024.110610

12. Morris ED, Ghanem AI, Dong M, Pantelic M V., Walker EM, Glide-Hurst CK. Cardiac substructure segmentation with deep learning for improved cardiac sparing. *Med Phys*. 2020;47(2):576-586. doi:10.1002/mp.13940
13. Summerfield N, Morris E, Banerjee S, et al. Enhancing Precision in Cardiac Segmentation for Magnetic Resonance-Guided Radiation Therapy Through Deep Learning. *International Journal of Radiation Oncology*Biology*Physics*. 2024;120(3):904-914. doi:10.1016/j.ijrobp.2024.05.013
14. Kataria T, Bisht SS, Gupta D, et al. Quantification of coronary artery motion and internal risk volume from ECG gated radiotherapy planning scans. *Radiotherapy and Oncology*. 2016;121(1):59-63. doi:10.1016/j.radonc.2016.08.006
15. Lester SC, Taparra K, Petersen MM, et al. Electrocardiogram-Gated Computed Tomography with Coronary Angiography for Cardiac Substructure Delineation and Sparing in Patients with Mediastinal Lymphomas Treated with Radiation Therapy. *Pract Radiat Oncol*. 2020;10(2):104-111. doi:10.1016/j.prro.2019.10.016
16. Lopez-Mattei J, Yang EH, Baldassarre LA, et al. Cardiac computed tomographic imaging in cardio-oncology: An expert consensus document of the Society of Cardiovascular Computed Tomography (SCCT). Endorsed by the International Cardio-Oncology Society (ICOS). *J Cardiovasc Comput Tomogr*. 2023;17(1):66-83. doi:10.1016/j.jcct.2022.09.002
17. He Q, Dong M, Summerfield N, Glide-Hurst C. MAGNET: A Modality-Agnostic Network for 3d Medical Image Segmentation. In: *2023 IEEE 20th International Symposium on Biomedical Imaging (ISBI)*. IEEE; 2023:1-5. doi:10.1109/ISBI53787.2023.10230587
18. He Q, Summerfield N, Dong M, Glide-Hurst C. Modality-Agnostic Learning for Medical Image Segmentation Using Multi-Modality Self-Distillation. In: *2024 IEEE International Symposium on Biomedical Imaging (ISBI)*. IEEE; 2024:1-5. doi:10.1109/ISBI56570.2024.10635881
19. Feng M, Moran JM, Koelling T, et al. Development and validation of a heart atlas to study cardiac exposure to radiation following treatment for breast cancer. *Int J Radiat Oncol Biol Phys*. 2011;79(1):10-18. doi:10.1016/j.ijrobp.2009.10.058
20. Duane F, Aznar MC, Bartlett F, et al. A cardiac contouring atlas for radiotherapy. *Radiotherapy and Oncology*. 2017;122(3):416-422. doi:10.1016/j.radonc.2017.01.008
21. Socha J, Rygielska A, Uzięblo-Życzkowska B, et al. Contouring cardiac substructures on average intensity projection 4D-CT for lung cancer radiotherapy: A proposal of a heart valve contouring atlas. *Radiotherapy and Oncology*. 2022;167:261-268. doi:10.1016/j.radonc.2021.12.041
22. Loap P, Servois V, Dhonneur G, Kirov K, Fourquet A, Kirova Y. A Radiation Therapy Contouring Atlas for Cardiac Conduction Node Delineation. *Pract Radiat Oncol*. 2021;11(4):e434-e437. doi:10.1016/j.prro.2021.02.002
23. Sherer M V., Lin D, Elguindi S, et al. Metrics to evaluate the performance of auto-segmentation for radiation treatment planning: A critical review. *Radiotherapy and Oncology*. 2021;160:185-191. doi:10.1016/J.RADONC.2021.05.003

24. Gupta S, Hu X, Kaan J, et al. Learning Topological Interactions for Multi-Class Medical Image Segmentation. In: 2022:701-718. doi:10.1007/978-3-031-19818-2_40
25. Byrne N, Clough JR, Montana G, King AP. A Persistent Homology-Based Topological Loss Function for Multi-class CNN Segmentation of Cardiac MRI. In: 2021:3-13. doi:10.1007/978-3-030-68107-4_1
26. Isensee F, Kickingereder P, Wick W, Bendszus M, Maier-Hein KH. No New-Net. In: 2019:234-244. doi:10.1007/978-3-030-11726-9_21
27. Zhao Y, Wang X, Che T, Bao G, Li S. Multi-task deep learning for medical image computing and analysis: A review. *Comput Biol Med*. 2023;153:106496. doi:10.1016/j.compbiomed.2022.106496
28. Pan LS, Li CW, Su SF, Tay SY, Tran QV, Chan WP. Coronary artery segmentation under class imbalance using a U-Net based architecture on computed tomography angiography images. *Sci Rep*. 2021;11(1):14493. doi:10.1038/s41598-021-93889-z
29. Hurkmans C, Bibault JE, Brock KK, et al. A joint ESTRO and AAPM guideline for development, clinical validation and reporting of artificial intelligence models in radiation therapy. *Radiotherapy and Oncology*. 2024;197:110345. doi:10.1016/j.radonc.2024.110345
30. Rong Y, Chen Q, Fu Y, et al. NRG Oncology Assessment of Artificial Intelligence Deep Learning–Based Auto-segmentation for Radiation Therapy: Current Developments, Clinical Considerations, and Future Directions. *International Journal of Radiation Oncology*Biology*Physics*. 2024;119(1):261-280. doi:10.1016/j.ijrobp.2023.10.033
31. Zhuang X, Li L, Payer C, et al. Evaluation of algorithms for Multi-Modality Whole Heart Segmentation: An open-access grand challenge. *Med Image Anal*. 2019;58. doi:10.1016/j.media.2019.101537
32. Havaei M, Guizard N, Chapados N, Bengio Y. HeMIS: Hetero-Modal Image Segmentation. In: 2016:469-477. doi:10.1007/978-3-319-46723-8_54
33. Ding Y, Yu X, Yang Y. RFNet: Region-aware Fusion Network for Incomplete Multi-modal Brain Tumor Segmentation. In: *2021 IEEE/CVF International Conference on Computer Vision (ICCV)*. IEEE; 2021:3955-3964. doi:10.1109/ICCV48922.2021.00394
34. Zhang Y, He N, Yang J, et al. mmFormer: Multimodal Medical Transformer for Incomplete Multimodal Learning of Brain Tumor Segmentation. In: 2022:107-117. doi:10.1007/978-3-031-16443-9_11
35. Dou Q, Liu Q, Heng PA, Glocker B. Unpaired Multi-Modal Segmentation via Knowledge Distillation. *IEEE Trans Med Imaging*. 2020;39(7):2415-2425. doi:10.1109/TMI.2019.2963882
36. Banerjee S, Summerfield N, Dong M, Glide-Hurst C. Volumetric medical image segmentation through dual self-distillation in U-shaped networks. *IEEE Trans Biomed Eng*. Published online 2025:1-14. doi:10.1109/TBME.2025.3566995
37. Wang J, Yang Y, Mao J, Huang Z, Huang C, Xu W. CNN-RNN: A Unified Framework for Multi-label Image Classification. In: *2016 IEEE Conference on Computer Vision and Pattern Recognition (CVPR)*. IEEE; 2016:2285-2294. doi:10.1109/CVPR.2016.251
38. Sutskever I, Martens J, Dahl G, Hinton G. On the importance of initialization and momentum in deep learning. In: Dasgupta S, McAllester D, eds. *Proceedings of the*

*30th International Conference on Machine Learning*. Vol 28. Proceedings of Machine Learning Research. PMLR; 2013:1139-1147. https://proceedings.mlr.press/v28/sutskever13.html

39. Isensee F, Jaeger PF, Kohl SAA, Petersen J, Maier-Hein KH. nnU-Net: a self-configuring method for deep learning-based biomedical image segmentation. *Nat Methods*. 2021;18(2):203-211. doi:10.1038/s41592-020-01008-z
40. Paszke A, Gross S, Massa F, et al. PyTorch: An Imperative Style, High-Performance Deep Learning Library. In: Wallach H, Larochelle H, Beygelzimer A, d Alché-Buc F, Fox E, Garnett R, eds. *Advances in Neural Information Processing Systems*. Vol 32. Curran Associates, Inc.; 2019. https://proceedings.neurips.cc/paper_files/paper/2019/file/bdbca288fee7f92f2bfa9f7012727740-Paper.pdf
41. Cardoso MJ, Li W, Brown R, et al. MONAI: An open-source framework for deep learning in healthcare. Published online November 4, 2022.
42. Zhang C, Bengio S, Hardt M, Recht B, Vinyals O. Understanding deep learning (still) requires rethinking generalization. *Commun ACM*. 2021;64(3):107-115. doi:10.1145/3446776
43. Zhao R, Qian B, Zhang X, et al. Rethinking Dice Loss for Medical Image Segmentation. In: *2020 IEEE International Conference on Data Mining (ICDM)*. IEEE; 2020:851-860. doi:10.1109/ICDM50108.2020.00094
44. Specktor-Fadida B, Yehuda B, Link-Sourani D, Ben-Sira L, Ben-Bashat D, Joskowicz L. Contour Dice Loss for Structures with Fuzzy and Complex Boundaries in Fetal MRI. In: 2023:355-368. doi:10.1007/978-3-031-25066-8_19
45. Maier-Hein L, Reinke A, Godau P, et al. Metrics reloaded: recommendations for image analysis validation. *Nat Methods*. 2024;21(2):195-212. doi:10.1038/s41592-023-02151-z
46. Shit S, Paetzold JC, Sekuboyina A, et al. clDice - a Novel Topology-Preserving Loss Function for Tubular Structure Segmentation. In: *2021 IEEE/CVF Conference on Computer Vision and Pattern Recognition (CVPR)*. IEEE; 2021:16555-16564. doi:10.1109/CVPR46437.2021.01629
47. Wasserthal J, Breit HC, Meyer MT, et al. TotalSegmentator: Robust Segmentation of 104 Anatomic Structures in CT Images. *Radiol Artif Intell*. 2023;5(5). doi:10.1148/ryai.230024
48. Gharleghi R, Adikari D, Ellenberger K, et al. Automated segmentation of normal and diseased coronary arteries – The ASOCA challenge. *Computerized Medical Imaging and Graphics*. 2022;97:102049. doi:10.1016/j.compmedimag.2022.102049
49. Nordstrom M, Hult H, Löfman F, Söderberg J. On Image Segmentation With Noisy Labels: Characterization and Volume Properties of the Optimal Solutions to Accuracy and Dice. In: Koyejo S, Mohamed S, Agarwal A, Belgrave D, Cho K, Oh A, eds. *Advances in Neural Information Processing Systems*. Vol 35. Curran Associates, Inc.; 2022:34321-34333. https://proceedings.neurips.cc/paper_files/paper/2022/file/ddf6eeeaa92957d3100b217a4428d819-Paper-Conference.pdf

50. Loap P, De Marzi L, Kirov K, et al. Development of Simplified Auto-Segmentable Functional Cardiac Atlas. *Pract Radiat Oncol*. 2022;12(6):533-538. doi:10.1016/j.prro.2022.02.004
51. Finnegan RN, Chin V, Chlap P, et al. Open-source, fully-automated hybrid cardiac substructure segmentation: development and optimisation. *Phys Eng Sci Med*. 2023;46(1):377-393. doi:10.1007/s13246-023-01231-w
52. Warner E, Lee J, Hsu W, et al. Multimodal Machine Learning in Image-Based and Clinical Biomedicine: Survey and Prospects. *Int J Comput Vis*. 2024;132(9):3753-3769. doi:10.1007/s11263-024-02032-8
53. Podobnik G, Ibragimov B, Tappeiner E, et al. HaN-Seg: The head and neck organ-at-risk CT and MR segmentation challenge. *Radiotherapy and Oncology*. 2024;198:110410. doi:10.1016/j.radonc.2024.110410
54. Lempart M, Nilsson MP, Scherman J, et al. Pelvic U-Net: multi-label semantic segmentation of pelvic organs at risk for radiation therapy anal cancer patients using a deeply supervised shuffle attention convolutional neural network. *Radiation Oncology*. 2022;17(1):114. doi:10.1186/s13014-022-02088-1

## Appendix A – Methods

### Imaging Pre-Processing

To improve model generalizability, all clinical images underwent similar preprocessing steps. Images were resampled to 1.5 x 1.5 x 1.5 $mm^3$ resolution and semi-automatically cropped to the WH with manual verification that the full cardiac anatomy was included. For validation and testing cohorts, all images underwent similar center-cropping for model input through a sliding window inference across the cropped volume. With application to clinical use cases, an in-house heart detection software is utilized to identify the volume of interest with manual verification. For the MMWHS dataset, all images were re-oriented to a common axis and underwent similar pre-processing steps as the clinical dataset.

CT-based images were windowed to (-250, 550) HU following institutional protocol for cardiology image review. All images then underwent an instance-based Z-score normalization (subtraction by the image mean, scaled by the image standard deviation) across the cropped image.

### Fractionated Images

Fractionated images were included as training augmentation for the clinical dataset as described in prior work[13] for specifically MR-Linac images when training MAGIC. The fractionated images are acquired across multiple days of treatment for a given patient, representing new patient anatomy due to intra-fraction differences. Yet, fraction images still represent the same patient. As such, these volumes are leveraged as complimentary datasets during training but cannot be considered unique volumes for use in validation or testing. By utilizing these volumes in training, the MR-Linac cohort was expanded to match the Sim-CT and CCTA cohorts where patient data was more abundant. Although these images are not randomly generated per iteration like typical training augmentations are, they are referred to as such in this work as they're solely used in training to further diversify the information provided to the model. For each patient in the MR-Linac training cohort, their simulation (day 0) scan is used alongside 2-3 fraction images, better matching the Sim-CT and CCTA cohorts. For both validation and testing, only unique patients were included to maximize the representation of population anatomy.

### Training Augmentations

Additional training augmentations were applied to each image for each modality prior to input during fitting to improve generalizability. Augmentations include random flipping along each axis; rotations of both 90° and between -10° and 10° along each axis; shifting and scaling pixel intensity between -10% and 10%; and randomly cropping the images to smaller input patches of 96 x 96 x 96 voxels.

### Optimization

Models were optimized for 500 epochs using an AdamW optimizer (learning rate, $10^{-3}$; weight decay, $10^{-5}$) and a poly learning rate reduction (gamma, 0.9). All modality-specific encoders and group specific decoders of MAGIC are connected to a single optimizer. MAGIC is implemented and optimized in PyTorch. When the loss function is calculated for backpropagation, only the parameters contributing to final prediction are updated when stepping the optimizer. Meaning, only the modality's specific encoder changes when presented with an input and each segmentation group's decoder remains able to specialize in their task. As such, a single optimizer is capable of fitting MAGIC to multiple input modalities and segmentation tasks, conserving momentum for the shared bottleneck and decoder branches that update every iteration.

Each decoder in MAGIC is individually targeted with a composite loss function tailored to the segmentation task of that branch. The composite loss is the summation of a Dice, Dice-boundary (following Specktor-Fadida et al.[44]), and focal loss. For the MMWHS experiment and segmentation groups A and B of the clinical experiment, targeting the whole heart, chambers, and great vessels, a focal loss with a gamma of 0 is used, resulting in cross-entropy. For groups C and D of the clinical experiment, targeting the coronary arteries, valves, and conduction nodes, a gamma of 2 is used to aid the greater class imbalance present between the smaller structures and the background.

Magic on the clinical dataset was trained in two steps to leverage semi-supervised training. First, MAGIC underwent fully supervised training on manually delineated reference volumes. Once trained, this initial model was used to predict 20 substructures on unlabeled cases from the same institutes. Each prediction was visually approved for accuracy against the underlying anatomy by a trained cardiac anatomy delineator. No edits were performed on the predicted images. In total, an additional 75 images (25 images per modality) with predicted labels were included in a second round of semi-supervised training for a newly initialized MAGIC following the same training protocol and hyper-parameters as the first. These same images were used when training all comparator, unimodal models for equivalent comparisons.

**Decoder Ablation**

MAGIC handles overlapping structures by treating each group of non-overlapping structures as a separate task in multi-task learning. By splitting the decoder at a certain point and utilizing multi-head or multi-branch layers, a specific output can be assigned to each task. In this work, a 6-layer nnU-Net forms the backbone structure where the 6$^{th}$ layer represents the bottleneck. The point along the decoder for the remaining 5 layers where the layers are split into task-specific branches is a hyperparameter that requires ablating to better understand the effect on the final segmentation performance.

To find the optimal point along the decoding branch to split the modules into multi-branch parallel layers, we empirically ablate the decoder design on the clinical dataset as shown in Table A1 on the clinical dataset. Split levels encompass a single, combined output after the final layer, a multi-head layer, and parallel multi-branch decoders, 2-5 layers deep. Splitting the decoder into parallel branches following the bottleneck output demonstrated the best performance over all three modalities as well as the best improvement for MR-Linac DSC (e.g. the lowest performing modality).

**MAGIC Decoder Ablation**

| Split Level | | Sim-CT | | MR-Linac | | CCTA | | All MAGIC | |
|---|---|---|---|---|---|---|---|---|---|
| | | DSC (AU) | HD95 (mm) | DSC (AU) | HD95 (mm) | DSC (AU) | HD95 (mm) | DSC (AU) | HD95 (mm) |
| Combined Output | | **0.760 ± 0.154** | **5.13 ± 2.62** | 0.664 ± 0.223 | 7.48 ± 4.77 | 0.804 ± 0.154 | 5.12 ± 3.40 | 0.743 ± 0.189 | 5.91 ± 3.87 |
| Multi-head layer | | 0.759 ± 0.151 | 5.31 ± 2.75 | 0.667 ± 0.221 | 7.72 ± 5.79 | 0.802 ± 0.155 | 5.31 ± 3.91 | 0.743 ± 0.187 | 6.11 ± 4.48 |
| Multi-Branch | Last 2 layers | 0.755 ± 0.164 | 5.62 ± 5.88 | 0.670 ± 0.219 | **7.16 ± 4.64** | 0.804 ± 0.157 | 5.16 ± 3.54 | 0.743 ± 0.190 | 5.98 ± 4.86 |
| | Last 3 layers | 0.760 ± 0.157 | 5.18 ± 2.65 | 0.668 ± 0.222 | 7.30 ± 4.75 | 0.801 ± 0.158 | 5.34 ± 4.00 | 0.743 ± 0.190 | 5.94 ± 4.02 |
| | Last 4 layers | 0.746 ± 0.172 | 5.33 ± 2.66 | 0.672 ± 0.222 | 7.35 ± 5.76 | 0.803 ± 0.161 | 5.18 ± 3.55 | 0.740 ± 0.194 | 5.95 ± 3.88 |
| | All layers | 0.753 ± 0.163 | 5.29 ± 2.78 | **0.678 ± 0.209** | 7.31 ± 4.84 | **0.805 ± 0.157** | **5.09 ± 3.58** | **0.745 ± 0.185** | **5.90 ± 3.96** |

Table A1: MAGIC ablation empirically comparing split positions to introduce the multi-branch parallel decoder. Abbreviations: AU, Arbitrary units; defined in text.

## Encoder Selection

MAGIC's modality-specific encoders require manual selection depending on the input being used. During training, the inputs are known and provided at input to guide encoder selection. In a real use case, this information can be extracted from an image's information header, conserving autonomous use.

## Appendix B – Results

## MMWHS Performance Results

**MMWHS Evaluation**

| | | DSC (AU) | | | NSD (AU; τ = 3 mm) | | | HD95 (mm) | | | MASD (mm) | | |
|---|---|---|---|---|---|---|---|---|---|---|---|---|---|
| | | MAGIC | Unimodal | p<0.05 | MAGIC | Unimodal | p<0.05 | MAGIC | Unimodal | p<0.05 | MAGIC | Unimodal | p<0.05 |
| CCTA | Average | 0.88 ± 0.08 | 0.85 ± 0.13 | * | 0.91 ± 0.09 | 0.85 ± 0.15 | * | 8.9 ± 12.0 | 11.7 ± 13.4 | * | 2.2 ± 2.4 | 3.2 ± 4.9 | * |
| | Chambers | 0.87 ± 0.09 | 0.83 ± 0.14 | * | 0.90 ± 0.10 | 0.82 ± 0.16 | * | 6.4 ± 7.9 | 11.9 ± 13.7 | * | 1.9 ± 1.8 | 3.6 ± 5.6 | * |
| | Myo | 0.84 ± 0.08 | 0.80 ± 0.08 | | 0.96 ± 0.03 | 0.91 ± 0.03 | | 3.3 ± 1.1 | 4.8 ± 0.5 | | 1.2 ± 0.3 | 1.5 ± 0.2 | |
| | LA | 0.82 ± 0.14 | 0.71 ± 0.24 | | 0.83 ± 0.13 | 0.66 ± 0.27 | | 9.6 ± 7.7 | 20.7 ± 22.4 | | 2.0 ± 0.9 | 8.6 ± 10.7 | |
| | LV | 0.94 ± 0.02 | 0.91 ± 0.01 | | 0.97 ± 0.03 | 0.90 ± 0.03 | | 3.1 ± 0.8 | 6.0 ± 1.6 | | 1.2 ± 0.4 | 1.6 ± 0.4 | |
| | RA | 0.86 ± 0.07 | 0.85 ± 0.05 | | 0.87 ± 0.08 | 0.82 ± 0.06 | | 12.1 ± 13.3 | 11.1 ± 9.9 | | 3.4 ± 3.3 | 2.9 ± 2.1 | |
| | RV | 0.90 ± 0.02 | 0.87 ± 0.05 | | 0.90 ± 0.08 | 0.81 ± 0.08 | | 4.2 ± 1.2 | 17.0 ± 11.9 | | 1.6 ± 0.3 | 3.2 ± 1.8 | |
| | GVs | 0.91 ± 0.07 | 0.89 ± 0.07 | | 0.91 ± 0.09 | 0.91 ± 0.08 | | 15.1 ± 17.1 | 11.2 ± 12.9 | | 3.1 ± 3.4 | 2.1 ± 2.4 | |
| | AA | 0.95 ± 0.03 | 0.96 ± 0.01 | | 0.96 ± 0.06 | 0.97 ± 0.02 | | 12.0 ± 18.2 | 2.7 ± 1.3 | | 2.5 ± 3.6 | 0.6 ± 0.2 | |
| | PA | 0.86 ± 0.06 | 0.83 ± 0.05 | | 0.86 ± 0.08 | 0.85 ± 0.08 | | 18.3 ± 15.3 | 19.6 ± 13.7 | | 3.8 ± 3.1 | 3.6 ± 2.6 | |
| MR | Average | 0.87 ± 0.04 | 0.84 ± 0.07 | * | 0.89 ± 0.06 | 0.86 ± 0.09 | * | 11.0 ± 14.3 | 13.4 ± 16.4 | * | 1.7 ± 1.1 | 2.1 ± 2.0 | * |
| | Chambers | 0.88 ± 0.04 | 0.86 ± 0.07 | * | 0.90 ± 0.06 | 0.87 ± 0.09 | * | 5.4 ± 2.1 | 8.4 ± 7.9 | * | 1.5 ± 0.5 | 2.0 ± 1.9 | * |
| | Myo | 0.83 ± 0.03 | 0.79 ± 0.06 | | 0.94 ± 0.03 | 0.90 ± 0.08 | | 3.9 ± 0.9 | 7.1 ± 5.3 | | 1.2 ± 0.2 | 1.3 ± 0.2 | |
| | LA | 0.87 ± 0.03 | 0.86 ± 0.03 | | 0.86 ± 0.07 | 0.85 ± 0.09 | | 6.6 ± 1.3 | 7.9 ± 2.4 | | 1.7 ± 0.7 | 1.8 ± 0.8 | |
| | LV | 0.92 ± 0.03 | 0.91 ± 0.03 | | 0.95 ± 0.02 | 0.92 ± 0.05 | | 3.9 ± 0.5 | 5.4 ± 3.1 | | 1.2 ± 0.3 | 1.4 ± 0.4 | |
| | RA | 0.88 ± 0.01 | 0.87 ± 0.01 | | 0.87 ± 0.04 | 0.86 ± 0.04 | | 7.0 ± 2.6 | 7.5 ± 2.2 | | 1.9 ± 0.5 | 2.0 ± 0.4 | |
| | RV | 0.89 ± 0.04 | 0.84 ± 0.10 | | 0.90 ± 0.06 | 0.84 ± 0.14 | | 5.6 ± 2.2 | 14.1 ± 14.9 | | 1.5 ± 0.5 | 3.5 ± 3.7 | |
| | GVs | 0.83 ± 0.03 | 0.81 ± 0.05 | | 0.86 ± 0.05 | 0.84 ± 0.07 | | 24.9 ± 20.8 | 26.1 ± 23.7 | | 2.4 ± 1.7 | 2.5 ± 2.2 | |
| | AA | 0.81 ± 0.03 | 0.80 ± 0.05 | | 0.82 ± 0.03 | 0.81 ± 0.08 | | 30.2 ± 20.0 | 30.6 ± 22.1 | | 2.0 ± 1.4 | 1.6 ± 0.5 | |
| | PA | 0.85 ± 0.02 | 0.83 ± 0.03 | | 0.90 ± 0.03 | 0.87 ± 0.04 | | 19.6 ± 20.2 | 21.6 ± 24.4 | | 2.7 ± 1.9 | 3.3 ± 2.8 | |
| Combined | Average | 0.87 ± 0.07 | 0.85 ± 0.10 | * | 0.90 ± 0.08 | 0.85 ± 0.12 | * | 9.9 ± 13.2 | 12.6 ± 15.0 | * | 2.0 ± 1.9 | 2.6 ± 3.8 | * |
| | Chambers | 0.88 ± 0.07 | 0.84 ± 0.11 | * | 0.90 ± 0.08 | 0.85 ± 0.13 | * | 5.9 ± 5.8 | 10.2 ± 11.3 | * | 1.7 ± 1.3 | 2.8 ± 4.3 | * |
| | Myo | 0.84 ± 0.06 | 0.79 ± 0.07 | * | 0.95 ± 0.03 | 0.91 ± 0.06 | * | 3.6 ± 1.0 | 5.9 ± 3.9 | * | 1.2 ± 0.3 | 1.4 ± 0.2 | * |
| | LA | 0.85 ± 0.10 | 0.78 ± 0.19 | * | 0.84 ± 0.11 | 0.75 ± 0.22 | * | 8.1 ± 5.8 | 14.3 ± 17.2 | * | 1.8 ± 0.8 | 5.2 ± 8.3 | * |
| | LV | 0.93 ± 0.02 | 0.91 ± 0.03 | * | 0.96 ± 0.03 | 0.91 ± 0.04 | * | 3.5 ± 0.8 | 5.7 ± 2.4 | | 1.2 ± 0.4 | 1.5 ± 0.4 | * |
| | RA | 0.87 ± 0.05 | 0.86 ± 0.04 | | 0.87 ± 0.07 | 0.84 ± 0.06 | * | 9.5 ± 9.9 | 9.3 ± 7.4 | | 2.6 ± 2.5 | 2.4 ± 1.6 | |
| | RV | 0.90 ± 0.03 | 0.86 ± 0.08 | * | 0.90 ± 0.07 | 0.83 ± 0.12 | * | 4.9 ± 1.9 | 15.5 ± 13.6 | * | 1.5 ± 0.4 | 3.4 ± 2.9 | * |
| | GVs | 0.87 ± 0.06 | 0.85 ± 0.07 | | 0.89 ± 0.07 | 0.88 ± 0.08 | | 20.0 ± 19.6 | 18.6 ± 20.5 | | 2.8 ± 2.7 | 2.3 ± 2.3 | |
| | AA | 0.88 ± 0.08 | 0.88 ± 0.09 | | 0.89 ± 0.08 | 0.89 ± 0.10 | | 21.1 ± 21.2 | 16.7 ± 21.0 | | 2.3 ± 2.7 | 1.1 ± 0.6 | |
| | PA | 0.85 ± 0.04 | 0.83 ± 0.05 | * | 0.88 ± 0.06 | 0.86 ± 0.06 | * | 18.9 ± 17.9 | 20.6 ± 19.8 | | 3.3 ± 2.6 | 3.5 ± 2.7 | |

Table B1: Quantitative performance of MAGIC and unimodal comparator models on the Multi-Modality Whole-Heart Segmentation (MMWHS) challenge dataset[31] for Dice similarity coefficient (DSC), normalized surface distance (NSD; tolerance, 3mm), 95% Hausdorff distance, (HD95) and mean absolute surface distance (MASD) on CCTA (top), MR (middle), and all combined (bottom) inputs. Abbreviations: Arbitrary units, AU; rest defined in text.

## Training History

Training history including composite loss values and validation DSC for the clinical experiment are presented in Figure B1. MAGIC fits all three modalities and all four segmentation groups at the same time, resulting in a single history profile. Meanwhile, the unimodal comparator models have separate histories per modality and per group.

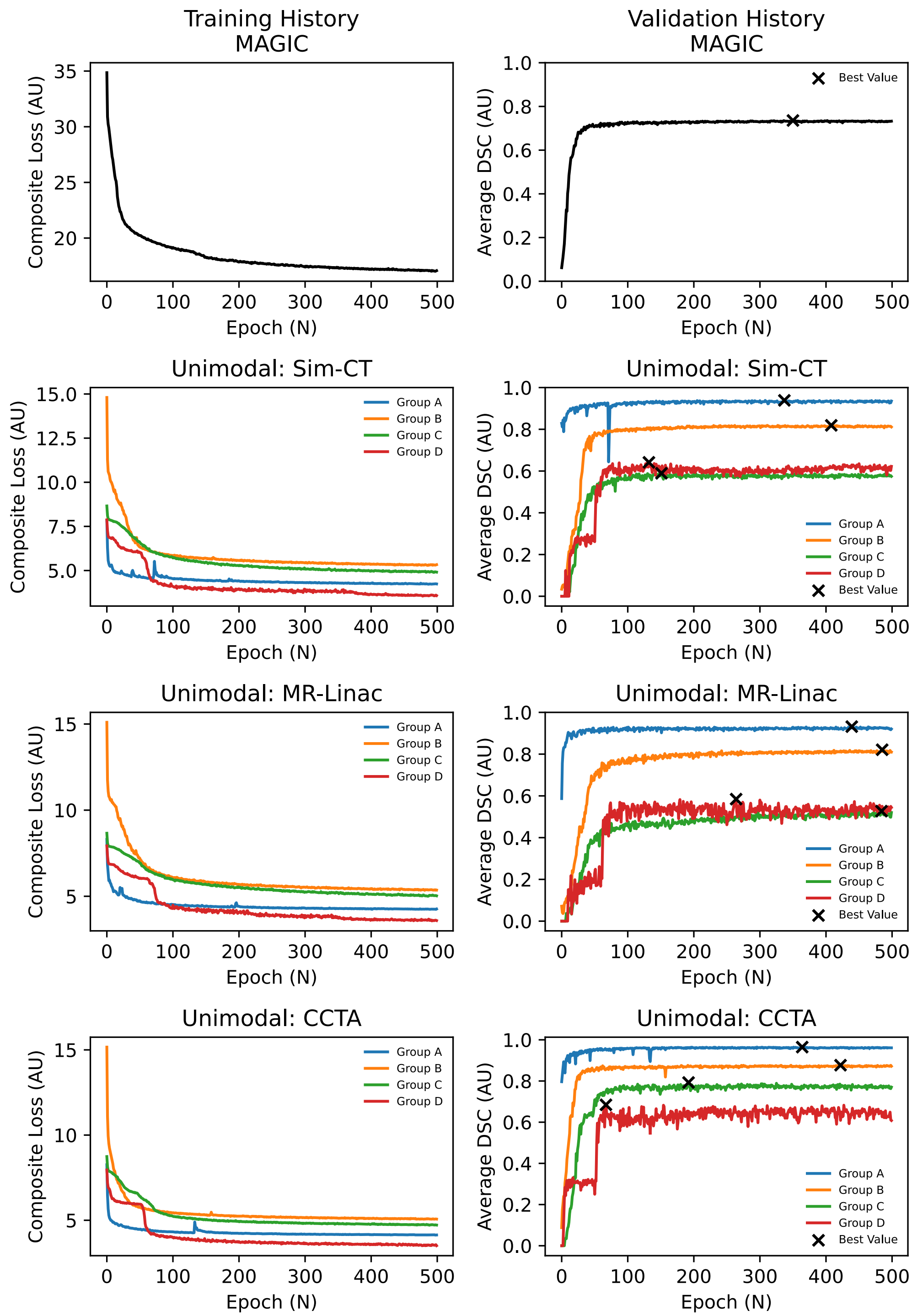

Figure B1: Training loss (Left) and hold-out validation (Right) history for MAGIC and unimodal comparison models on the clinical dataset. Unimodal groups are as follows: A) whole heart; B) chambers and great vessels; C) coronary arteries and valves; D) conduction nodes. Abbreviations: Arbitrary Units, AU; defined in text.

## Sim-CT Performance Metrics

**Sim-CT Quantitative Comparison**

| | DSC (AU) | | | NSD (AU; τ = 3 mm) | | | HD95 (mm) | | | MASD (mm) | | |
|---|---|---|---|---|---|---|---|---|---|---|---|---|
| | MAGIC | Unimodal | p<0.05 | MAGIC | Unimodal | p<0.05 | MAGIC | Unimodal | p<0.05 | MAGIC | Unimodal | p<0.05 |
| Composite | 0.75 ± 0.16 | 0.75 ± 0.16 | | 0.87 ± 0.13 | 0.87 ± 0.14 | | 5.3 ± 2.8 | 5.6 ± 3.5 | | 1.7 ± 0.8 | 1.7 ± 0.9 | |
| Whole Heart | 0.96 ± 0.01 | 0.96 ± 0.01 | | 0.93 ± 0.05 | 0.94 ± 0.04 | | 4.3 ± 1.5 | 4.3 ± 1.4 | | 1.3 ± 0.4 | 1.3 ± 0.3 | |
| Chambers | 0.89 ± 0.05 | 0.89 ± 0.04 | | 0.89 ± 0.09 | 0.89 ± 0.08 | | 5.8 ± 2.9 | 5.5 ± 2.3 | | 1.6 ± 0.6 | 1.6 ± 0.4 | |
| RA | 0.87 ± 0.06 | 0.88 ± 0.04 | | 0.88 ± 0.11 | 0.89 ± 0.08 | | 5.7 ± 2.5 | 5.1 ± 1.6 | | 1.8 ± 0.9 | 1.7 ± 0.5 | |
| LA | 0.88 ± 0.03 | 0.87 ± 0.03 | * | 0.89 ± 0.06 | 0.88 ± 0.06 | | 5.2 ± 1.5 | 5.5 ± 1.3 | | 1.6 ± 0.4 | 1.7 ± 0.4 | * |
| RV | 0.88 ± 0.05 | 0.89 ± 0.04 | * | 0.87 ± 0.11 | 0.89 ± 0.09 | | 7.1 ± 4.3 | 6.1 ± 3.2 | | 1.5 ± 0.4 | 1.4 ± 0.3 | * |
| LV | 0.93 ± 0.02 | 0.93 ± 0.02 | | 0.90 ± 0.08 | 0.91 ± 0.06 | | 5.1 ± 2.4 | 5.2 ± 2.5 | | 1.4 ± 0.4 | 1.4 ± 0.3 | |
| Great Vessels | 0.81 ± 0.09 | 0.81 ± 0.09 | | 0.90 ± 0.10 | 0.89 ± 0.11 | | 5.3 ± 2.8 | 5.4 ± 3.1 | | 1.5 ± 0.7 | 1.5 ± 0.6 | |
| AA | 0.88 ± 0.06 | 0.88 ± 0.05 | | 0.90 ± 0.10 | 0.91 ± 0.09 | | 4.8 ± 2.8 | 4.4 ± 2.2 | | 1.4 ± 0.6 | 1.4 ± 0.5 | |
| SVC | 0.80 ± 0.08 | 0.79 ± 0.07 | | 0.93 ± 0.07 | 0.92 ± 0.08 | | 4.8 ± 3.1 | 5.0 ± 2.8 | | 1.6 ± 0.9 | 1.7 ± 0.8 | |
| IVC | 0.75 ± 0.11 | 0.73 ± 0.13 | | 0.79 ± 0.14 | 0.78 ± 0.15 | | 7.3 ± 3.2 | 8.4 ± 4.4 | | 2.0 ± 1.0 | 1.7 ± 0.8 | |
| PA | 0.86 ± 0.04 | 0.86 ± 0.04 | | 0.94 ± 0.05 | 0.94 ± 0.05 | | 4.2 ± 1.9 | 4.2 ± 1.7 | | 1.3 ± 0.5 | 1.3 ± 0.4 | |
| PV | 0.76 ± 0.04 | 0.76 ± 0.05 | | 0.93 ± 0.03 | 0.93 ± 0.04 | | 5.1 ± 1.4 | 4.9 ± 1.6 | | 1.3 ± 0.4 | 1.4 ± 0.3 | |
| Coronary Arteries | 0.60 ± 0.13 | 0.61 ± 0.11 | | 0.88 ± 0.11 | 0.89 ± 0.09 | | 6.0 ± 3.2 | 6.9 ± 4.7 | | 1.7 ± 0.7 | 1.6 ± 0.6 | |
| LMCA | 0.64 ± 0.10 | 0.66 ± 0.08 | | 0.92 ± 0.09 | 0.94 ± 0.07 | | 3.7 ± 0.9 | 4.1 ± 1.5 | | 1.4 ± 0.4 | 1.3 ± 0.2 | |
| LADA | 0.59 ± 0.13 | 0.60 ± 0.14 | | 0.86 ± 0.10 | 0.85 ± 0.12 | | 8.5 ± 4.1 | 12.3 ± 6.2 | | 1.8 ± 0.7 | 1.6 ± 0.8 | |
| RCA | 0.57 ± 0.16 | 0.60 ± 0.11 | | 0.84 ± 0.13 | 0.87 ± 0.08 | | 6.4 ± 2.7 | 5.8 ± 1.9 | | 1.7 ± 0.8 | 1.7 ± 0.6 | |
| LCx | 0.58 ± 0.10 | 0.59 ± 0.10 | | 0.89 ± 0.08 | 0.88 ± 0.08 | | 5.2 ± 1.8 | 5.3 ± 1.5 | | 1.7 ± 0.5 | 1.8 ± 0.5 | |
| Valves | 0.70 ± 0.18 | 0.69 ± 0.19 | | 0.85 ± 0.18 | 0.83 ± 0.21 | * | 4.7 ± 2.6 | 5.5 ± 4.2 | * | 1.9 ± 1.1 | 2.0 ± 1.3 | |
| V-AV | 0.71 ± 0.16 | 0.69 ± 0.19 | | 0.83 ± 0.20 | 0.76 ± 0.25 | | 3.8 ± 1.8 | 4.3 ± 1.9 | | 2.0 ± 1.2 | 2.1 ± 1.3 | |
| V-PV | 0.79 ± 0.17 | 0.77 ± 0.17 | | 0.91 ± 0.17 | 0.91 ± 0.18 | | 3.7 ± 2.1 | 3.8 ± 2.2 | | 1.5 ± 1.1 | 1.6 ± 1.2 | |
| V-TV | 0.67 ± 0.16 | 0.67 ± 0.14 | | 0.85 ± 0.15 | 0.83 ± 0.16 | | 5.7 ± 2.9 | 6.9 ± 5.5 | | 2.0 ± 1.0 | 2.3 ± 1.5 | * |
| V-MV | 0.63 ± 0.20 | 0.61 ± 0.23 | | 0.83 ± 0.16 | 0.80 ± 0.20 | * | 5.7 ± 2.8 | 7.1 ± 4.8 | | 2.1 ± 1.0 | 1.9 ± 1.0 | |
| Nodes | 0.66 ± 0.12 | 0.65 ± 0.15 | | 0.79 ± 0.19 | 0.77 ± 0.20 | | 4.6 ± 1.7 | 5.0 ± 2.0 | | 2.2 ± 0.8 | 2.3 ± 1.1 | |
| N-SA | 0.65 ± 0.10 | 0.61 ± 0.10 | * | 0.76 ± 0.16 | 0.72 ± 0.16 | | 4.8 ± 1.4 | 5.4 ± 1.3 | * | 2.2 ± 0.7 | 2.5 ± 0.7 | |
| N-AV | 0.67 ± 0.14 | 0.69 ± 0.17 | | 0.82 ± 0.21 | 0.81 ± 0.23 | | 4.4 ± 1.8 | 4.5 ± 2.5 | | 2.1 ± 1.0 | 2.1 ± 1.3 | |

Table B2: Quantitative analysis comparing MAGIC and unimodal models across 20 substructures for Sim-CT inputs. Abbreviations: Arbitrary Units, AU; defined in text.

## MR-Linac Performance Metrics

**MR-Linac Quantitative Comparison**

| | DSC (AU) | | | NSD (AU; τ = 3 mm) | | | HD95 (mm) | | | MASD (mm) | | |
|---|---|---|---|---|---|---|---|---|---|---|---|---|
| | MAGIC | Unimodal | p<0.05 | MAGIC | Unimodal | p<0.05 | MAGIC | Unimodal | p<0.05 | MAGIC | Unimodal | p<0.05 |
| Composite | 0.68 ± 0.21 | 0.67 ± 0.23 | | 0.77 ± 0.18 | 0.76 ± 0.21 | | 7.3 ± 4.8 | 9.4 ± 23.6 | | 2.3 ± 1.2 | 4.1 ± 23.4 | |
| Whole Heart | 0.94 ± 0.01 | 0.94 ± 0.01 | | 0.88 ± 0.06 | 0.87 ± 0.05 | | 5.3 ± 1.1 | 5.7 ± 1.4 | | 1.8 ± 0.4 | 1.9 ± 0.4 | |
| Chambers | 0.87 ± 0.05 | 0.87 ± 0.05 | | 0.84 ± 0.11 | 0.83 ± 0.11 | * | 5.8 ± 2.3 | 6.0 ± 2.3 | | 1.9 ± 0.6 | 2.0 ± 0.7 | * |
| RA | 0.84 ± 0.05 | 0.83 ± 0.05 | | 0.79 ± 0.12 | 0.76 ± 0.11 | * | 6.9 ± 2.7 | 7.4 ± 2.8 | | 2.3 ± 0.7 | 2.5 ± 0.7 | * |
| LA | 0.87 ± 0.03 | 0.86 ± 0.04 | | 0.85 ± 0.07 | 0.84 ± 0.09 | | 6.2 ± 2.2 | 6.2 ± 2.2 | | 1.8 ± 0.4 | 2.0 ± 0.5 | |
| RV | 0.86 ± 0.05 | 0.86 ± 0.05 | | 0.82 ± 0.11 | 0.82 ± 0.12 | | 6.0 ± 2.1 | 6.0 ± 2.0 | | 2.1 ± 0.6 | 2.1 ± 0.7 | |
| LV | 0.93 ± 0.01 | 0.92 ± 0.02 | | 0.92 ± 0.05 | 0.91 ± 0.06 | * | 4.3 ± 1.2 | 4.4 ± 1.1 | | 1.5 ± 0.3 | 1.6 ± 0.4 | |
| Great Vessels | 0.72 ± 0.18 | 0.70 ± 0.20 | | 0.79 ± 0.18 | 0.78 ± 0.20 | | 9.2 ± 7.8 | 9.2 ± 7.1 | | 2.1 ± 1.2 | 2.1 ± 1.1 | |
| AA | 0.84 ± 0.05 | 0.84 ± 0.05 | | 0.83 ± 0.10 | 0.85 ± 0.10 | | 6.1 ± 2.7 | 5.9 ± 2.4 | | 2.1 ± 0.9 | 2.1 ± 0.7 | |
| SVC | 0.76 ± 0.16 | 0.71 ± 0.23 | * | 0.88 ± 0.16 | 0.83 ± 0.21 | * | 4.8 ± 3.0 | 6.5 ± 4.8 | * | 1.6 ± 0.8 | 1.9 ± 1.1 | |
| IVC | 0.57 ± 0.24 | 0.55 ± 0.27 | | 0.61 ± 0.25 | 0.61 ± 0.28 | | 13.4 ± 9.0 | 13.0 ± 9.3 | | 3.0 ± 1.9 | 2.7 ± 1.8 | |
| PA | 0.81 ± 0.07 | 0.81 ± 0.07 | | 0.84 ± 0.13 | 0.84 ± 0.14 | | 7.4 ± 6.3 | 7.5 ± 4.5 | | 1.9 ± 0.7 | 1.9 ± 0.8 | |
| PV | 0.62 ± 0.09 | 0.62 ± 0.09 | | 0.77 ± 0.10 | 0.76 ± 0.11 | | 14.4 ± 9.6 | 13.1 ± 8.0 | | 1.9 ± 0.6 | 2.2 ± 0.7 | |
| Coronary Arteries | 0.50 ± 0.18 | 0.50 ± 0.20 | | 0.76 ± 0.19 | 0.76 ± 0.21 | | 7.6 ± 3.6 | 8.4 ± 5.4 | | 2.4 ± 1.5 | 2.6 ± 1.9 | |
| LMCA | 0.51 ± 0.22 | 0.52 ± 0.24 | | 0.75 ± 0.28 | 0.76 ± 0.30 | | 6.3 ± 4.0 | 6.8 ± 4.5 | | 2.4 ± 2.4 | 2.5 ± 2.5 | |
| LADA | 0.61 ± 0.10 | 0.60 ± 0.11 | | 0.85 ± 0.08 | 0.84 ± 0.10 | | 6.7 ± 1.7 | 9.6 ± 7.7 | | 2.0 ± 0.4 | 2.2 ± 1.1 | |
| RCA | 0.46 ± 0.20 | 0.46 ± 0.23 | | 0.72 ± 0.20 | 0.71 ± 0.25 | | 9.3 ± 4.9 | 8.8 ± 5.2 | | 2.8 ± 1.5 | 3.3 ± 2.5 | |
| LCx | 0.41 ± 0.10 | 0.42 ± 0.12 | | 0.73 ± 0.08 | 0.73 ± 0.10 | | 8.0 ± 1.7 | 8.5 ± 2.0 | | 2.4 ± 0.6 | 2.4 ± 0.6 | |
| Valves | 0.62 ± 0.16 | 0.62 ± 0.17 | | 0.75 ± 0.17 | 0.75 ± 0.19 | | 6.8 ± 3.4 | 6.8 ± 3.4 | | 2.5 ± 1.1 | 2.5 ± 1.2 | |
| V-AV | 0.66 ± 0.10 | 0.64 ± 0.12 | | 0.79 ± 0.12 | 0.77 ± 0.21 | | 5.9 ± 1.4 | 5.3 ± 2.1 | | 2.0 ± 0.7 | 2.4 ± 0.8 | |
| V-PV | 0.66 ± 0.15 | 0.68 ± 0.17 | | 0.76 ± 0.21 | 0.77 ± 0.21 | | 7.2 ± 5.7 | 7.2 ± 5.1 | | 2.4 ± 1.0 | 2.1 ± 1.1 | |
| V-TV | 0.53 ± 0.21 | 0.51 ± 0.20 | | 0.68 ± 0.20 | 0.67 ± 0.18 | | 7.5 ± 2.7 | 7.9 ± 2.7 | | 3.1 ± 1.5 | 3.3 ± 1.4 | |
| V-MV | 0.64 ± 0.10 | 0.63 ± 0.12 | | 0.79 ± 0.10 | 0.78 ± 0.11 | | 6.6 ± 1.3 | 6.9 ± 2.1 | | 2.3 ± 0.8 | 2.4 ± 0.9 | |
| Nodes | 0.52 ± 0.16 | 0.45 ± 0.27 | | 0.60 ± 0.20 | 0.54 ± 0.32 | | 7.0 ± 2.4 | 25.6 ± 70.9 | | 3.0 ± 1.1 | 20.6 ± 71.7 | |
| N-SA | 0.52 ± 0.14 | 0.49 ± 0.27 | | 0.60 ± 0.18 | 0.61 ± 0.32 | | 7.0 ± 2.1 | 9.7 ± 9.6 | | 3.1 ± 1.1 | 4.8 ± 5.3 | |
| N-AV | 0.52 ± 0.18 | 0.41 ± 0.26 | | 0.59 ± 0.22 | 0.48 ± 0.30 | | 7.0 ± 2.6 | 41.4 ± 97.2 | | 2.8 ± 1.0 | 36.3 ± 98.7 | |

Table B3: Quantitative analysis comparing MAGIC and unimodal models across 20 substructures for MR-Linac inputs. The comparator model failed to predict the N-AV on a test case, resulting in infinity for both HD95 and MASD. For statistical comparison, we performed imputation, substituting this value with the image diagonal following published recommendations[45] and performed a Mann-Whitney U-Test following previous work[13]. Abbreviations: Arbitrary Units, AU; defined in text.

## CCTA Performance Metrics

**CCTA Quantitative Comparison**

| | DSC (AU) | | | NSD (AU; $\tau$ = 3 mm) | | | HD95 (mm) | | | MASD (mm) | | |
|---|---|---|---|---|---|---|---|---|---|---|---|---|
| | MAGIC | Unimodal | p<0.05 | MAGIC | Unimodal | p<0.05 | MAGIC | Unimodal | p<0.05 | MAGIC | Unimodal | p<0.05 |
| Composite | 0.80 ± 0.16 | 0.80 ± 0.16 | | 0.90 ± 0.14 | 0.90 ± 0.14 | | 5.1 ± 3.6 | 5.3 ± 3.7 | | 1.4 ± 1.2 | 1.4 ± 1.3 | |
| Whole Heart | 0.95 ± 0.01 | 0.95 ± 0.01 | | 0.88 ± 0.06 | 0.87 ± 0.04 | | 6.1 ± 2.7 | 6.4 ± 2.7 | | 1.7 ± 0.4 | 1.7 ± 0.3 | |
| Chambers | 0.93 ± 0.04 | 0.93 ± 0.04 | * | 0.95 ± 0.05 | 0.96 ± 0.05 | * | 4.2 ± 2.6 | 4.2 ± 2.8 | | 1.2 ± 0.4 | 1.2 ± 0.5 | |
| RA | 0.93 ± 0.01 | 0.93 ± 0.01 | * | 0.95 ± 0.03 | 0.95 ± 0.02 | * | 4.0 ± 1.0 | 3.9 ± 0.8 | | 1.2 ± 0.2 | 1.1 ± 0.2 | * |
| LA | 0.95 ± 0.01 | 0.95 ± 0.01 | * | 0.97 ± 0.02 | 0.97 ± 0.02 | | 3.4 ± 1.3 | 3.4 ± 1.6 | | 1.0 ± 0.2 | 0.9 ± 0.2 | |
| RV | 0.87 ± 0.04 | 0.88 ± 0.04 | | 0.90 ± 0.06 | 0.91 ± 0.06 | | 6.7 ± 3.7 | 7.0 ± 4.0 | | 1.8 ± 0.5 | 1.8 ± 0.5 | |
| LV | 0.96 ± 0.01 | 0.96 ± 0.01 | | 0.98 ± 0.01 | 0.99 ± 0.01 | * | 2.6 ± 0.5 | 2.5 ± 0.5 | | 0.9 ± 0.1 | 0.9 ± 0.1 | |
| Great Vessels | 0.84 ± 0.06 | 0.85 ± 0.06 | * | 0.90 ± 0.06 | 0.91 ± 0.06 | | 6.0 ± 3.4 | 6.0 ± 3.7 | | 1.3 ± 0.6 | 1.2 ± 0.5 | * |
| AA | 0.93 ± 0.02 | 0.93 ± 0.02 | * | 0.94 ± 0.05 | 0.95 ± 0.05 | | 4.5 ± 2.0 | 4.0 ± 1.8 | | 1.1 ± 0.4 | 1.0 ± 0.4 | * |
| SVC | 0.85 ± 0.04 | 0.86 ± 0.03 | | 0.92 ± 0.06 | 0.93 ± 0.05 | | 4.9 ± 2.2 | 4.7 ± 1.4 | | 1.2 ± 0.5 | 1.1 ± 0.4 | |
| IVC | 0.82 ± 0.05 | 0.82 ± 0.04 | | 0.91 ± 0.06 | 0.91 ± 0.05 | | 5.0 ± 2.0 | 4.7 ± 1.1 | | 1.5 ± 0.7 | 1.5 ± 0.5 | |
| PA | 0.84 ± 0.05 | 0.84 ± 0.05 | | 0.87 ± 0.08 | 0.87 ± 0.08 | | 8.1 ± 5.3 | 8.8 ± 6.2 | | 1.6 ± 0.6 | 1.4 ± 0.5 | * |
| PV | 0.78 ± 0.03 | 0.79 ± 0.04 | * | 0.88 ± 0.02 | 0.88 ± 0.03 | | 7.8 ± 2.0 | 7.6 ± 2.3 | | 1.0 ± 0.2 | 1.0 ± 0.2 | * |
| Coronary Arteries | 0.77 ± 0.12 | 0.77 ± 0.13 | | 0.94 ± 0.07 | 0.94 ± 0.07 | | 5.4 ± 4.7 | 5.6 ± 4.4 | | 0.8 ± 0.5 | 0.8 ± 0.5 | |
| LMCA | 0.73 ± 0.07 | 0.72 ± 0.10 | | 0.95 ± 0.07 | 0.94 ± 0.07 | | 4.1 ± 2.5 | 4.6 ± 2.6 | | 1.0 ± 0.4 | 1.0 ± 0.4 | |
| LADA | 0.80 ± 0.06 | 0.80 ± 0.06 | | 0.96 ± 0.04 | 0.95 ± 0.04 | | 4.3 ± 2.4 | 3.9 ± 2.1 | | 0.8 ± 0.3 | 0.7 ± 0.3 | |
| RCA | 0.86 ± 0.04 | 0.87 ± 0.03 | * | 0.96 ± 0.03 | 0.97 ± 0.03 | | 4.9 ± 3.9 | 4.9 ± 3.8 | | 0.4 ± 0.1 | 0.4 ± 0.1 | |
| LCx | 0.67 ± 0.16 | 0.67 ± 0.16 | | 0.90 ± 0.09 | 0.89 ± 0.09 | | 8.4 ± 7.0 | 8.9 ± 6.0 | | 1.1 ± 0.7 | 1.1 ± 0.8 | |
| Valves | 0.68 ± 0.23 | 0.68 ± 0.24 | | 0.84 ± 0.25 | 0.83 ± 0.25 | | 4.9 ± 3.9 | 5.3 ± 4.1 | * | 2.1 ± 2.2 | 2.2 ± 2.4 | |
| V-AV | 0.77 ± 0.13 | 0.78 ± 0.11 | | 0.95 ± 0.09 | 0.93 ± 0.06 | | 3.0 ± 1.0 | 3.6 ± 1.3 | | 1.4 ± 0.7 | 1.5 ± 0.8 | |
| V-PV | 0.38 ± 0.25 | 0.35 ± 0.23 | | 0.50 ± 0.26 | 0.46 ± 0.24 | | 9.7 ± 5.1 | 10.6 ± 5.0 | * | 4.5 ± 3.3 | 5.0 ± 3.4 | * |
| V-TV | 0.73 ± 0.11 | 0.77 ± 0.08 | * | 0.92 ± 0.11 | 0.94 ± 0.08 | * | 3.7 ± 1.2 | 3.8 ± 1.5 | | 1.6 ± 0.5 | 1.4 ± 0.4 | * |
| V-MV | 0.84 ± 0.04 | 0.83 ± 0.05 | | 0.98 ± 0.01 | 0.98 ± 0.02 | | 3.0 ± 0.7 | 3.1 ± 0.8 | | 0.9 ± 0.2 | 0.9 ± 0.2 | |
| Nodes | 0.72 ± 0.11 | 0.69 ± 0.12 | | 0.86 ± 0.16 | 0.83 ± 0.17 | | 3.8 ± 1.5 | 4.6 ± 2.0 | * | 1.9 ± 0.7 | 2.2 ± 1.0 | |
| N-SA | 0.71 ± 0.12 | 0.66 ± 0.14 | * | 0.86 ± 0.17 | 0.77 ± 0.20 | * | 3.9 ± 1.5 | 5.3 ± 2.4 | * | 1.9 ± 0.8 | 2.6 ± 1.1 | * |
| N-AV | 0.73 ± 0.11 | 0.72 ± 0.07 | | 0.87 ± 0.14 | 0.88 ± 0.09 | | 3.7 ± 1.5 | 3.8 ± 1.0 | | 1.9 ± 0.7 | 1.8 ± 0.5 | |

Table B4: Quantitative analysis comparing MAGIC and unimodal models across 20 substructures for CCTA inputs. Abbreviations: Arbitrary Units, AU; defined in text.

## Comparison Against TotalSegmentator

MAGIC's performance on Sim-CT volumes was compared to the performance of TotalSegmentator[47] tasks for the heart chambers, high resolution, and coronary arteries. Due to differences in label definitions between TotalSegmentator and the presented clinical dataset, the predicted myocardium and left ventricle of TotalSegmentator were combined into a single structure and the predicted aorta was truncated through morphological overlap of the reference aorta and a 6mm isotropic expansion to remove the predicted aortic branch and descending aorta. Similarly, MAGIC's and the reference coronary arteries were combined into a single structure to match the output of TotalSegmentator coronary artery task. Results on the Sim-CT test cohort are presented in Table B5. On Sim-CT, TotalSegmentator predicted the chambers, pulmonary artery, and ascending aorta with confidence, rivaling the performance of MAGIC on these structures despite differences in label definitions. In contrast, TotalSegmentator failed to produce anything useable for the coronary arteries (DSC of 0.01).

**Sim-CT**

| | | DSC (AU) | | HD95 (mm) | | MASD (mm) | |
|---|---|---|---|---|---|---|---|
| | | MAGIC | Total Segmentator | MAGIC | Total Segmentator | MAGIC | Total Segmentator |
| Task 1 | RA | 0.87 ± 0.06 | 0.82 ± 0.05 | 5.7 ± 2.5 | 8.0 ± 2.8 | 1.8 ± 0.9 | 2.3 ± 0.6 |
| | LA | 0.88 ± 0.03 | 0.69 ± 0.07 | 5.2 ± 1.5 | 39.2 ± 24.8 | 1.6 ± 0.4 | 5.8 ± 9.2 |
| | RV | 0.88 ± 0.05 | 0.81 ± 0.06 | 7.1 ± 4.3 | 21.8 ± 35.3 | 1.5 ± 0.4 | 5.6 ± 7,8 |
| | LV | 0.93 ± 0.02 | 0.82 ± 0.14 | 5.1 ± 2.4 | 33.6 ± 50.4 | 1.4 ± 0.4 | 15.1 ± 25.6 |
| | AA | 0.88 ± 0.06 | 0.80 ± 0.05 | 4.8 ± 2.8 | 8.5 ± 2.7 | 1.4 ± 0.6 | 2.3 ± 0.4 |
| | PA | 0.86 ± 0.04 | 0.75 ± 0.04 | 4.2 ± 1.9 | 10.6 ± 3.9 | 1.3 ± 0.5 | 2.6 ± 0.7 |
| Task 2 | Coronary Arteries | 0.62 ± 0.10 | 0.01 ± 0.01 | 8.3 ± 3.8 | 46.4 ± 40.9 | 1.7 ± 0.5 | 23.2 ± 12.7 |

Table B5: Comparison between MAGIC and TotalSegmentator[47] performance on the same Sim-CT test set for TotalSegmentator[47] task 1 (heart chambers and great vessels, top) and task 2 (coronary arteries, single classification, bottom). Abbreviations: Arbitrary Units, AU; defined in text.